\documentclass[11pt,a4paper]{article}
\pdfoutput=1
\usepackage{jheppub}
\usepackage{comment}
\usepackage{amsmath}
\usepackage{amssymb}
\usepackage{graphicx,color,slashed,braket}
\usepackage{bbm}
\usepackage{ifpdf}
\usepackage{float}
\usepackage{hhline}
\usepackage{soul}
\usepackage{bm}
\usepackage{tikz}  
\usetikzlibrary{arrows,shapes,trees,positioning} 
\usepackage{adjustbox}

\usepackage{empheq}
\def\beq{\begin{equation}}
\def\eq{\end{equation}}
\def\bea{\begin{eqnarray}}
\def\eea{\end{eqnarray}}

\newcommand{\be}{\begin{equation}} 
\newcommand{\ee}{\end{equation}} 
\newcommand{\barr}{\begin{array}}
\newcommand{\earr}{\end{array}}
\newcommand{\ba}{\begin{eqnarray}}
\newcommand{\ea}{\end{eqnarray}}

\newcommand{\lp}{\left(}
\newcommand{\rp}{\right)}

\tikzset{
  square arrow/.style={
    to path={-- ++(-10,-.25) -| (\tikztotarget)}
  }
}

\def\rmi{{\rm i}}

\def\K{{K\"{a}hler} }

\begin{document}

% preprint number: LAPTH-037/23

\title{On Asymptotic Dark Energy in String Theory}

\author[1,2]{Sera Cremonini,}
\emailAdd{cremonini@lehigh.edu}
\author[1]{Eduardo Gonzalo,}
\emailAdd{eduardo.gonzalo@lehigh.edu}
\author[1]{Muthusamy Rajaguru,}
\emailAdd{muthusamy.rajaguru@lehigh.edu}
\author[1]{Yuezhang Tang,}
\emailAdd{yut318@lehigh.edu}
\author[1,3]{and Timm Wrase}
\emailAdd{timm.wrase@lehigh.edu}

\affiliation[1]{Department of Physics, Lehigh University, Bethlehem, PA, 18018, USA}
\affiliation[2]{
Kavli Institute of Theoretical Physics, University of California Santa Barbara, Santa Barbara,CA, 93106, USA}
\affiliation[3]{
Laboratoire d’Annecy-le-Vieux de Physique Th\'eorique (LAPTh),
CNRS, Universit\'e Savoie Mont Blanc (USMB), 
%9 Chemin de Bellevue, 
74940 Annecy, France}

\abstract{We examine bounds on accelerated expansion in asymptotic regions of the moduli space in string theory compactifications to four spacetime dimensions. While there are conjectures that forbid or constrain accelerated expansion in such asymptotic regions, potential counter examples have been discussed recently in the literature. We check whether such counter examples can arise in explicit string theory constructions, focusing in particular on non-geometric compactifications of type IIB string theory that have no \K moduli. We find no violation of the Strong Asymptotic dS Conjecture and thus provide support for the absence of accelerated expansion in asymptotic regions of a barely explored corner of the string landscape. 
Moreover, working in a simplified setting, we point out a new mechanism for potentially connecting 
the Sharpened Distance Conjecture and the Strong Asymptotic dS Conjecture. If this argument could be generalized, it would mean that the Sharpened Distance Conjecture is implied by the Strong Asymptotic dS Conjecture, and that their exponential factors are naturally related by a factor of 2.
}

%\notoc
\maketitle

\section{Introduction}

One of the major challenges in string phenomenology is the construction of universes with positive vacuum energy that explain the observed dark energy \cite{SupernovaSearchTeam:1998fmf, SupernovaCosmologyProject:1996grv}. Current cosmological data is compatible both with a constant and a time dependent vacuum energy \cite{Planck:2018vyg}. The first possibility means our universe is described by either a stable or metastable de Sitter (dS) vacuum. From the model building point of view this requires the stabilization of all moduli in a non-supersymmetric vacuum. Without supersymmetry it is more difficult to guarantee control over the low energy four-dimensional effective action governing the theory. However, multiple efforts have been directed towards finding dS vacua in string theory, starting with the seminal papers \cite{Kachru:2003aw,  Balasubramanian:2005zx}. These constructions stabilize moduli at finite values and argue that the corrections that are neglected do not change the results. Given that not all such corrections are explicitly known, there is an on-going discussion about the validity of all the existing dS constructions in string theory, see for example \cite{Danielsson:2018ztv, Gao:2020xqh, Lust:2022lfc, Junghans:2022exo, Cicoli:2023opf}. The second scenario consists in building a so-called \textit{quintessence} model, where scalar fields are allowed to roll slowly \cite{Hellerman:2001yi, Fischler:2001yj, Kaloper:2008qs, Cicoli:2012tz, Cicoli:2018kdo, Hebecker:2019csg, ValeixoBento:2020ujr, Brinkmann:2022oxy}. From a model building point of view this option needs to meet additional requirements; for example, there are bounds on the size of the mass of the quintessence fields.

Before aspiring to build realistic quintessence models from string theory, it is more prudent to establish whether the latter allows for quintessence at all. Asymptotic limits in moduli space have the advantage of providing more control due to suppressed corrections and are easier to study. As a first step, it would therefore be useful to understand if scalar potentials arising in string theory could asymptotically allow for quintessence. Such a question fits perfectly into the theme of the swampland program \cite{Vafa:2005ui, vanBeest:2021lhn, Agmon:2022thq, Palti:2019pca, Grana:2021zvf}, which aims to gain insights into the validity of low-energy effective field theories (EFTs) using quantum gravity. A common procedure within the swampland program is to collect sets of results and ideas and bundle them together in a single statement in the form of a conjecture. The asymptotic behavior of scalar potentials arising in string theory has been the subject of several swampland investigations \cite{Ooguri:2006in, Obied:2018sgi, Banlaki:2018ayh, Junghans:2018gdb, Ooguri:2018wrx, Hebecker:2018vxz, Lust:2019zwm} resulting in the swampland dS Conjecture discussed in detail in the next paragraph. In addition to this particular conjecture, there exist several lower bounds on the slow-roll parameters which were obtained in the context of searches for dS vacua and inflation in specific string theory setups \cite{Hertzberg:2007wc, Haque:2008jz, Flauger:2008ad, Caviezel:2009tu, Wrase:2010ew, Shiu:2011zt, Andriot:2016xvq, Garg:2018reu, Andriot:2018ept, Garg:2018zdg, Andriot:2022xjh, Andriot:2023isc}. These papers provide no-go theorems that apply  at any point in moduli space and usually forbid slow-roll inflation by showing that the slow-roll parameter $\epsilon_V$ characterizing the slope of the potential obeys $\epsilon_V \geq 1$. Hence they help narrow down the search for asymptotic accelerated expansion from slowly rolling scalar fields by excluding large classes of models. 

The scalar potentials arising from string compactifications in the asymptotic limits of moduli space are expected to take the following form, $ V \sim e^{-\gamma \phi}$, 
where $\phi$ is a canonical field and $\gamma$ is an $O(1)$ number. For accelerated expansion in four spacetime dimensions we would require $\gamma = \frac{|| \nabla V(\phi) ||}{V(\phi)} < \sqrt{2}$.  It was shown in \cite{Cicoli:2021fsd} that type IIA/B and heterotic Calabi-Yau compactifications cannot have $\gamma<\sqrt{2}$ in the large volume, weak coupling limit. The swampland dS Conjecture in the asymptotic limit \cite{Ooguri:2018wrx, Hebecker:2018vxz}, which is a generalization of the Dine-Seiberg problem \cite{Dine:1985he}, provides a rough bound of $\gamma \gtrsim \mathcal{O}(1)$ and hence provides preliminary motivation for the potential absence of asymptotic quintessence. Using dimensional reduction techniques \cite{Rudelius:2021azq, Rudelius:2021oaz, Etheredge:2022opl}, the bound was sharpened by Rudelius \cite{Rudelius:2022gbz}, see also \cite{Bedroya:2019snp}. This led to the so-called Strong Asymptotic dS Conjecture, which states that $\gamma \geq \sqrt{2}$ in four dimensions and more generically $\gamma \geq 2/\sqrt{d-2}$ in $d$ dimensions. This forbids in any asymptotic limit of string theory compactifications the existence of solutions that are undergoing accelerated expansion (see also \cite{Marconnet:2022fmx, 
Bedroya:2022tbh, Apers:2022cyl,  Shiu:2023nph, Shiu:2023rxt} for closely related work).\footnote{Accelerated expansion is forbidden for $\gamma\geq \sqrt{2}$, if the 4d spacetime has no curvature, which is the case we focus on in this paper. For negative curvature it was shown in \cite{Marconnet:2022fmx} that accelerated expansion is actually possible.}

The authors of \cite{Calderon-Infante:2022nxb} have pointed out a potential loophole that could then allow us to realize quintessence asymptotically in string theory. In particular, they argue that having two terms compete asymptotically in the scalar potential would provide a possible way out of the conjecture of \cite{Rudelius:2022gbz}, and they identify several potential examples that could violate the bound of \cite{Rudelius:2022gbz}. As already noted by the authors, the examples of \cite{Calderon-Infante:2022nxb}  do not account for the stabilization of the volume modulus. If the latter is not stabilized then this runaway direction is too steep to preserve accelerated expansion. The possible loophole suggested in \cite{Calderon-Infante:2022nxb}  is intriguing and we believe requires additional investigation.

In this work, we attempt to take the first steps towards a realization of the scenario described above, in which two terms in the potential compete asymptotically. We will focus on non-geometric type IIB models whose moduli space includes only the dilaton and the complex structure moduli. These models were first introduced in \cite{Becker:2006ks, Becker:2007dn} and were shown to be a good testing ground for swampland conjectures in \cite{Ishiguro:2021csu, Bardzell:2022jfh, Becker:2022hse}. We show that in our explicit model there is no violation of the Strong Asymptotic dS Conjecture. 
Thus, we are showing the absence of accelerated expansion in asymptotic limits of moduli space in a non-geometric and thus very stringy setup. 

We also recall the previous observation that the minimal exponential prefactors in the Strong Asymptotic dS Conjecture and the Sharpened Distance Conjecture \cite{Ooguri:2006in, Klaewer:2016kiy, Etheredge:2022opl} are related by a factor of 2. As mentioned above, for the Strong Asymptotic dS Conjecture we have $\gamma \geq \gamma_{min} = 2/\sqrt{d-2}$ in a generic $d$-dimensional spacetime. For the Sharpened Distance Conjecture a tower of massive states becomes light in the asymptotic limit $\phi \to \infty$, with masses of the form $m(\phi) \sim e^{-\alpha \phi}$ with $\alpha \geq \alpha_{min} = 1/\sqrt{d-2}$. Somewhat surprisingly $\gamma_{min}=2\alpha_{min}$. An argument was given in \cite{Etheredge:2022opl} for why the Sharpened Distance Conjecture, together with the Emergent String Conjecture \cite{Lee:2019wij, Lee:2019xtm}, implies the Strong Asymptotic dS Conjecture. In this paper, we provide a potential explanation for this factor of 2, which is distinct from the argument given in \cite{Etheredge:2022opl}. If our argument can be generalized, then this would mean that the Sharpened Distance Conjecture follows from the absence of asymptotic accelerated expansion in string theory. Thus, this would be complementary to the argument of \cite{Etheredge:2022opl}, and provide an interesting new motivation for the Sharpened Distance Conjecture.

The outline of this paper is as follows. In section \ref{review} we review previous results pertaining to slow-roll inflation and quintessence in string compactifications. In section \ref{sec:OurNewModel} we study the possibility of violating the Strong Asymptotic dS Conjecture in non-geometric compactifications of type IIB string theory. In section \ref{sec:DistanceStrongdS} we spell out evidence for a potential direct connection between the Strong Asymptotic dS Conjecture and the Sharpened Distance Conjecture. We summarize our results and conclude in section \ref{sec:Conclusions}. Two appendices contain technical details.

\section{Review of previous results} \label{review}
In this section, we review the status of obtaining accelerated expansion in asymptotic limits of the moduli space. We focus here on the case of slow-roll type inflation/quintessence in asymptotic limits of string theory compactifications to four dimensions. These setups are the most studied in the literature and provide an interesting test bed for the conjecture that accelerated expansion is forbidden in any asymptotic limit \cite{Rudelius:2022gbz}. In particular, it was argued in \cite{Calderon-Infante:2022nxb} that there could be counter-examples if one studies infinite distance limits in which not a single but multiple terms in the scalar potential are dominant in the asymptotic limit. The caveat in \cite{Calderon-Infante:2022nxb} was however that the authors studied only the complex structure sector of F-theory compactifications and did not stabilize the \K moduli. In these setups, there was thus always a steep direction associated with the overall volume of the CY$_4$ manifold that prevented any accelerated expansion. In the next section we study non-geometric type IIB flux compactifications \emph{without} \K moduli and in particular without a volume modulus \cite{Becker:2006ks, Becker:2007dn} to see whether the counter-examples of \cite{Calderon-Infante:2022nxb} can be realized in such setups. Before doing that we review in this section the previous related results in some detail.

Usually, the scalar potential falls off exponentially when we approach an asymptotic point in moduli space, i.e. $V(\phi) \sim e^{-\gamma \phi}$ for very large $\phi$ (see for example \cite{Ooguri:2018wrx, Hebecker:2018vxz} for a recent related discussion). While we find below cases where this is more subtle, it makes sense to phrase this discussion in terms of the prefactor $\gamma$. The requirement for slow-roll inflation, and a corresponding accelerated expansion, is that the two slow-roll parameters are smaller than one
\begin{equation}
\epsilon_V = \frac12 \lp \frac{|\nabla V|}{V}\rp^2 <1\,,  \qquad \eta_V = \left| \text{min}\lp \frac{\nabla_i \nabla_j V}{V}\rp \right| < 1\,.
\end{equation}
For a single scalar field with $V(\phi) \sim e^{-\gamma \phi}$ this amounts to the bound $\gamma<\sqrt{2}$ in four dimensions. On the other hand, the Strong Asymptotic dS Conjecture states that $\gamma \geq \sqrt{2}$ in asymptotic limits of the scalar field space, see \cite{Rudelius:2021azq, Rudelius:2021oaz, Etheredge:2022opl, Rudelius:2022gbz}.\footnote{The related Trans-Planckian Censorship Conjecture (TCC) \cite{Bedroya:2019snp} imposes the weaker bound $\gamma \geq \sqrt{2/3}$. However, notice that it was mentioned already in \cite{Bedroya:2019snp} that their bound can become stronger for exponentially decreasing potentials, implying the Strong Asymptotic dS Conjecture. For consistency with the literature we will reserve the name TCC for the weaker bound $\gamma \geq \sqrt{2/3}$.} Thus, the question of whether or not the Strong Asymptotic dS Conjecture is true is equivalent to the question of whether accelerated expansion is possible in asymptotic limits of the moduli space. Recently, the authors of \cite{Shiu:2023nph, Shiu:2023rxt} pointed out that slow-roll inflation/quintessence is non-generic since it requires $\epsilon_V = \gamma^2/2 \ll 1$, which is usually not the case. For a single field with an exponential potential $\epsilon = \epsilon_V = \gamma^2/2$ even if the slow-roll approximation is not valid. However, for more complicated trajectories in generic multifield potentials, one has to solve the full set of equations to check whether accelerated expansion is possible. This difference will not be important to us below since we will not find sufficiently shallow potentials.

Motivated by the observed accelerated expansion of our own universe \cite{SupernovaSearchTeam:1998fmf, SupernovaCosmologyProject:1996grv}, string theorists have been trying to construct corresponding models in string theory for more than two decades. This has led to many interesting new insights and also excluded certain classes of models that can never give rise to four dimensional spacetimes that undergo accelerated expansion. Next we recall the for us relevant no-go theorems and their loopholes.

In \cite{Hertzberg:2007wc} the authors proved the absence of dS vacua and slow-roll inflation in flux compactifications of massive type IIA on CY$_3$ manifolds. In order to do that they focused only on two universal moduli, the 4d dilaton $\tau$ and the overall volume modulus $\rho$. These scalar fields decouple from the other scalar fields in the sense that the slow-roll parameter $\epsilon_V$ takes the form \footnote{$\tau$ and $\rho$ are the commonly used scalar fields that are however not canonically normalized, leading to the additional factors in the expression for $\epsilon_V$.}
\begin{equation}\label{eq:epsilonrhotau}
    \epsilon_V = \frac13 \lp \frac{\rho \partial_\rho V}{V}\rp^2 +\frac14 \lp \frac{\tau \partial_\tau V}{V}\rp^2 + |\ldots|^2 \geq \frac13 \lp \frac{\rho \partial_\rho V}{V}\rp^2 +\frac14 \lp \frac{\tau \partial_\tau V}{V}\rp^2\,,
\end{equation}
where $|\ldots|^2$ denotes the positive or zero contribution from all the other scalar fields in the theory. Given the scaling of the scalar potential in type IIA Calabi-Yau flux compactifications, one can show that the two terms on the right-hand-side of equation \eqref{eq:epsilonrhotau} are larger or equal to 27/13 \cite{Hertzberg:2007wc}. This means that accelerated expansion is forbidden \emph{anywhere} in moduli space. Therefore the Strong Asymptotic dS Conjecture is trivially satisfied in massive type IIA flux compactifications on Calabi-Yau manifolds.

The above no-go theorem was extended to \emph{massless} type IIA compactifications on internal manifolds with curvature and many more refined no-go theorems were derived in \cite{Haque:2008jz, Flauger:2008ad, Caviezel:2009tu, Wrase:2010ew, Andriot:2016xvq, Andriot:2018ept}. While these no-go theorems aimed to excluded dS extrema, they usually prove a bound on $\epsilon_V$ that requires it to be larger than one anywhere in moduli space. This means in all such cases the Strong Asymptotic dS Conjecture holds. In cases where the no-goes allow for $\epsilon_V$ to be smaller than one or in cases where there is no no-go theorem at all, one can then ask the question of whether it is possible to have accelerated expansion in an asymptotic region.

The most important and easiest to study asymptotic regions would be a large volume and weak coupling regime in which both $\alpha'$ and string loop corrections are suppressed. This limit corresponds to sending $\rho$ and $\tau$ above to infinity and it ensures that one can fully trust the low-energy supergravity approximation to the particular string theory in question. Whether dS extrema could exist in this limit was studied in \cite{Junghans:2018gdb, Banlaki:2018ayh}, where it was shown that this is actually not possible in large classes of type IIA string theory compactifications. The authors of \cite{Cicoli:2021fsd} then asked whether quintessence could be realized with parametric control using the above form of $\epsilon_V$. They found that in the asymptotic large $\rho$ and $\tau$ limit of heterotic and type II string theory the potential is always too steep to allow for accelerated expansion from a slowly rolling scalar field \cite{Cicoli:2021fsd}. This motivated the question of whether any other asymptotic limit of string theory can give rise to an accelerating universe.

Rudelius \cite{Rudelius:2022gbz} noted in many examples that any limit in which one sends a single scalar field to zero leads to a scalar potential $V(\phi) \sim e^{-\gamma \phi}$ that is so steep that slow-roll inflation and therefore quintessence cannot arise. This then led to the Strong Asymptotic dS Conjecture that forbids such accelerated expansion in any asymptotic region of moduli space. However, as was noted in \cite{Calderon-Infante:2022nxb}, it is possible that two (or more) fields become large and approach an asymptotic limit such that two terms in the scalar potential compete with each other. For example, one could consider the limit $\phi_1= c\, \phi_2 \to \infty$ for $c \in \mathbb{R}_{>0}$ with the asymptotic scalar potential
\ba
    V(\phi_1,\phi_2) &=& A_1 e^{-\gamma_{1,1} \phi_1 -\gamma_{1,2} \phi_2 } +A_2 e^{-\gamma_{2,1} \phi_1 -\gamma_{2,2} \phi_2} +\ldots \cr 
    &=& A_1 e^{-(c \, \gamma_{1,1} +\gamma_{1,2}) \phi_2 } +A_2 e^{-(c \, \gamma_{2,1} +\gamma_{2,2}) \phi_2} +\ldots\,,
\ea
where $\ldots$ represents subleading terms. The first two terms above can scale exactly in the same way, i.e. $c\, \gamma_{1,1} +\gamma_{1,2} = c\, \gamma_{2,1} +\gamma_{2,2}$, but the individual scalings with respect to the moduli $\phi_1, \phi_2$ could be very different in the two terms. Such competing terms open up the possibility of getting accelerated expansion along the trajectory $\phi_1= c\, \phi_2 \to \infty$, even if two individual terms do not allow for this \cite{Calderon-Infante:2022nxb}. 
While this mechanism is  extremely interesting, the authors of \cite{Calderon-Infante:2022nxb} restricted themselves to the complex structure sector and did not study the stabilization of the \K moduli. The explicit potential had an overall $\rho^{-3}$ dependence, i.e., 
\be
    V(\phi_1,\phi_2,\rho) = \frac{1}{\rho^3} \lp A_1 e^{-(c\, \gamma_{1,1} +\gamma_{1,2}) \phi_2 } +A_2 e^{-(c\, \gamma_{2,1} +\gamma_{2,2}) \phi_2} +\ldots \rp\,.
\ee
From equation \eqref{eq:epsilonrhotau} above, we see that this leads to $\epsilon_V \geq 3$ and prohibits slow-roll inflation even if $\phi_1$ and $\phi_2$ contribute less than 1 to $\epsilon_V$. One could now delve into stabilizing the \K moduli while trying to stay in the asymptotic limits of the complex structure moduli space that were identified as potential counter-examples in \cite{Calderon-Infante:2022nxb}. However, given the lively debate related to this topic, we refrain from doing so and rather focus on string theory models without \K moduli \cite{Becker:2006ks, Becker:2007dn}. Using Landau-Ginzburg techniques it was shown to be possible to study flux compactifications on generalized Calabi-Yau manifolds with $h^{1,1}=0$ that are the mirror dual of rigid Calabi-Yau manifolds \cite{Becker:2006ks}. This removes the requirement of having to stabilize \K moduli in type IIB flux compactifications and leads to a variety of interesting results \cite{Becker:2006ks, Becker:2007dn, Ishiguro:2021csu, Bardzell:2022jfh, Becker:2022hse}. In the next section we will review the corresponding four-dimensional scalar field theories and try to exploit the loopholes pointed out in \cite{Calderon-Infante:2022nxb}. Before we do that we quickly review the results and some notation from \cite{Calderon-Infante:2022nxb} in the next subsection, so that we can build on these results in section \ref{sec:OurNewModel}.

\subsection{A possible loophole}
To find asymptotic trajectories that lead to accelerated expansion we must solve the Friedmann equations under the slow roll approximation. Consider the following 4-dimensional action  for scalar fields minimally coupled to gravity,
\be
    S=\int d^{4}x \sqrt{-g} \lp\frac{R}{2} - \frac{1}{2} M_{ab} \partial_{\mu} \phi^{a} \partial^{\mu} \phi^{b}-V(\phi)\rp,
\ee
where $M_{ab}$ is the real-valued metric on the scalar field space. The slow-roll Friedmann equations then amount to
\be
    \frac{d\phi^{a}}{dt} =-\frac{1}{\sqrt{3 V}} M^{ab} \partial_{b}V . \label{eqn:slowroll}
\ee
Asymptotic trajectories can be parametrized as $\phi^{a}(\sigma)$ as $\sigma \rightarrow \infty$. As pointed out in \cite{Calderon-Infante:2022nxb}, solving \eqref{eqn:slowroll} asymptotically can be simplified by taking ratios of pairs of these equations. 

In \cite{Calderon-Infante:2022nxb} the authors only study examples with two rolling fields and note that in order to find a unique asymptotic trajectory it is necessary to have two terms that  compete 
asymptotically in the potential. For a simple diagonal field space metric, we note here that a scalar potential with $n$ moving fields requires $n$ terms competing asymptotically to get a unique trajectory. Let us exemplify this by considering the following potential 
\be
    V = A_{1} \frac{U_{2}}{U_{1}U_{3}} + A_{2} \frac{1}{U_{2}}\,.
\ee
Such terms can arise in the scalar potential we discuss below in the next section. The corresponding diagonal field space metric in that case is
\begin{equation}\label{eq:Mmatrix}
	M^{-1} = \begin{pmatrix} 
	2 U_{1}^{2} & 0 &0  \\
	0 &  2 U_{2}^{2} & 0 \\
        0 & 0 & 2 U_{3}^{2}  \\
	\end{pmatrix}.
	\quad
\end{equation} 
We parameterize the trajectory by $U_{1} = \alpha_{1} \sigma^{\beta_{1}}, U_{2} = \alpha_{2} \sigma^{\beta_{2}}, U_{3} = \sigma$, where $\sigma \rightarrow \infty$. Then from \eqref{eqn:slowroll} we get,
\begin{align}
    \alpha_{1} \beta_{1} \sigma^{\beta_{1}-1} \frac{d\sigma}{dt} &= \frac{2}{\sqrt{3V}} \lp A_{1} \alpha_{2} \sigma^{\beta_{2}-1}\rp \nonumber\\
     \alpha_{2} \beta_{2} \sigma^{\beta_{2}-1} \frac{d\sigma}{dt} &=  \frac{2}{\sqrt{3V}} \lp A_{2} - 
    A_{1} \frac{\alpha_{2}^{2}}{\alpha_{1}} \sigma^{2\beta_{2}-\beta_{1}-1}\rp \\
    \frac{d\sigma}{dt} &= \frac{2}{\sqrt{3V}} A_{1} \frac{\alpha_{2}}{\alpha_{1}} \sigma^{\beta_{2}-\beta_{1}}.\nonumber
\end{align}
Next, following \cite{Calderon-Infante:2022nxb} we take ratios of pairs of the above equations and evaluate them in the limit $\sigma \rightarrow \infty$ to solve for $\alpha_{1}, \alpha_{2}$. The ratio of the first and the third equation gives $\beta_{1}=1$. The ratio of the second and third equations then gives,
\be \label{eqn:beta2}
     \beta_{2}  = \frac{A_{2} \sigma^{1-\beta_{2}} - 
    A_{1} \frac{\alpha_{2}^{2}}{\alpha_{1}} \sigma^{\beta_{2}-1}}{A_{1} \frac{\alpha_{2}^2}{\alpha_{1}} \sigma^{\beta_{2}-1}} \, .
\ee
For $\beta_{2}<1$ and $\sigma \rightarrow \infty$ the right-hand side evaluates to $\infty$, whereas for $\beta_{2} > 1$ and $\sigma \rightarrow \infty$ it evaluates to $-1$. Thus, $\beta_{2} = 1$ is the only solution to this equation in the limit $\sigma \rightarrow \infty$. Plugging $\beta_{1} = \beta_{2} = 1$ into the equation \eqref{eqn:beta2} we see that we can only determine one of $\alpha_{1}$ and $\alpha_{2}$. So, the path to infinity in the $(U_1,U_2,U_3)$ moduli space is not uniquely fixed. 

Another way  of seeing this is to define $\tilde{U} = U_{1} U_{3}$ and $\tilde{U}_{T} = \frac{U_{1}}{U_{3}}$. The field space metric still remains diagonal and is given by, 
\begin{equation}
	M^{-1} = \begin{pmatrix} 
	4 \tilde{U}^{2} & 0 &0  \\
	0 &  2 U_{2}^{2} & 0 \\
        0 & 0 & 4 \tilde{U}_{T}^{2}  \\
	\end{pmatrix}.
	\quad
\end{equation} 
The scalar potential in terms of the redefined fields is independent of $\tilde{U}_{T}$,
\be
V = A_{1} \frac{U_{2}}{\tilde{U}} + A_{2} \frac{1}{U_{2}}\,.
\ee
As a result of this and the diagonal field space metric, $\tilde{U}_{T}$ remains a flat direction and the ratio of $U_{1}, U_{3}$ is undetermined. Hence, the asymptotic trajectory is not uniquely fixed. This means in turn that all these trajectories have the same value of $\gamma$. 

In \cite{Calderon-Infante:2022nxb} it was noted that situations where the trajectory is not uniquely determined do not lead to $\gamma<\sqrt{2}$ when there are two rolling scalar fields. Although we do not find any obstruction to obtaining $\gamma<\sqrt{2}$ when only two terms compete with three rolling scalar fields, we find that in practice such scenarios do not arise in the model we study below. This is because either it is not possible to stabilize the dilaton and the axions or because it is not possible to make terms that blow up in the potential vanish without losing the competing terms that give rise to $\gamma<\sqrt{2}$.

While for simple diagonal field space metrics one needs $n$ competing terms for $n$ rolling scalar fields in order to have a unique trajectory, this is not the case anymore for (generic) non-diagonal field space metrics. We do not encounter those examples in our simple model below but it is not hard, for example, to write down a toy model that uniquely fixes the trajectory for three rolling fields  with only two terms in the scalar potential.

%\section{Type IIB models with $\mathbf{h^{1,1}=0}$}\label{sec:OurNewModel}
\section{Type IIB models with \texorpdfstring{$\mathbf{h^{1,1}=0}$}{h11=0}}\label{sec:OurNewModel}
In string theory it is possible to study compactifications on `non-geometric spaces' that have no K\"ahler moduli. In particular, in type IIB where it is not possible to stabilize K\"ahler moduli with RR- or NSNS-fluxes this provides a very interesting playground for studying truly stringy flux compactifications. A detailed description of the effective action for these IIB flux vacua arising from orientifolds of Landau-Ginzburg models was presented in \cite{Becker:2006ks, Becker:2007dn,Bardzell:2022jfh}. Even though these models do not have Kähler moduli, they do have many complex structure moduli. These could allow us to realize the loopholes pointed out in \cite{Calderon-Infante:2022nxb} in full-fledged models, without having to worry about K\"ahler moduli stabilization. In this section, we perform a systematic study of possible asymptotic limits and check whether they allow for accelerated expansion.

\subsection{Effective action}
We focus here on a specific orientifold studied in \cite{Becker:2006ks} and restrict to those moduli which are mirror to the untwisted sector of the Type IIA toroidal compactification. So, we restrict to a subset consisting of three universal complex structure moduli.\footnote{In principle one would also have to include twisted sector fields. In the mirror dual type IIA compactification these correspond to blow-up modes of orbifold singularities and it was argued in \cite{DeWolfe:2005uu} that these can in principle be stabilized at a different scale without disturbing the three untwisted moduli.}

The flux superpotential is given by the standard Gukov-Vafa-Witten formula \cite{Gukov:1999ya},
\be
     W = \int_{M} G_{3} \wedge \Omega,
\ee
where $G_{3} \equiv F_{3}-\tau H_{3}$, with the axio-dilaton $\tau=C_0 + \rmi\, e^{-\phi}$. When restricting to the untwisted sector this amounts to 
\ba \label{eqn:superpotential}
     W &=& (f^{0} - \tau h^{0}) U_{1} U_{2} U_{3} - (f^{1}-\tau h^{1}) U_{2} U_{3}-(f^{2}-\tau h^{2}) U_{1} U_{3}-(f^{3}-\tau h^{3}) U_{1} U_{2} \nonumber\\[2mm]
   &&+ (f_{1}-\tau h_{1}) U_{1}+(f_{2}-\tau h_{2}) U_{2}+(f_{3}-\tau h_{3}) U_{3} + (f_{0}- \tau h_{0})\,,  
\ea
where $U_{1},U_{2}$ and $U_{3}$ are the complex structure moduli in the untwisted sector. The K\"ahler potential is given by \cite{Becker:2007dn}
\be 
    K = -4 \log[-\rmi(\tau-\bar{\tau})] - \log[\rmi  \int_{M} \Omega \wedge \bar{\Omega} ]\,.
\ee
In terms of only the untwisted sector, we have
\be\label{eqn:Kahlerpotential}
    K = -4 \log[-\rmi(\tau - \bar{\tau})] - \log[\rmi(U_{1}-\bar{U}_{1})(U_{2}-\bar{U}_{2})(U_{3}-\bar{U}_{3})]\,. 
\ee
The factor of 4 does not arise in the dimensional reduction of 10-dimensional type IIB supergravity but can be understood by using mirror symmetry \cite{Becker:2007dn}. 

The tadpole cancellation condition is given by
\be
   \frac{1}{\tau-\bar{\tau}} \int_{M} G \wedge \bar{G} + N_{D3} = \frac{1}{2} N_{O3}\,.
\ee
For us, this translates to
\be\label{eq:tadpole}
    N_{flux} + N_{D3} =-h^{0} f_{0} + \sum_{i=1}^{3} f^{i} h_{i} - \sum_{i=1}^{3} f_{i} h^{i} + h_{0} f^{0} + N_{D3} = 12\,. 
\ee
Given all the above we can investigate if the scalar potential given by
\be
    V = e^{K} \left(K^{I\bar{J}} D_{I}W \overline{D_{J} W} - 3 |W|^{2}\right),
\ee
can lead to trajectories that support asymptotic accelerated expansion. 

The above model is mirror dual to a type IIA compatification with geometric and non-geometric fluxes. A particular such model was studied already in subsection 3.3.2. of \cite{Calderon-Infante:2022nxb}. That model corresponds in our notation to $s=\tau$, $u=U_{1}=U_{2}=U_{3}$ and it was shown to not give rise to asymptotic accelerated expansion. We will see below from a slightly different point of view that the case where all three $U_{j}$ are equal cannot give accelerated expansion. Thus, we will naturally be led to study the case where the $U_{j}$ scale differently.\footnote{We thank Irene Valenzuela for pointing out this important duality to us.}

In the following discussions, we will rescale the fluxes in the scalar potential such that we are at a point in field space where the axions are vanishing and the saxions that are not rolling are set equal to one. This can be achieved by temporarily giving up the quantization of the fluxes. This has been employed before in the literature to make the analysis of scalar potentials easier. One can calculate the scalar potential and its derivative at the point where all non-rolling saxions are set equal to 1 and all axions are set equal to zero. Then via rescaling of the fluxes, one can reach any other points in moduli space. However, generically the tadpole cancellation conditions in equation \eqref{eq:tadpole} prevent one from reaching a point of arbitrary weak coupling or arbitrary large complex structure. We show the details of this rescaling for our model in appendix \ref{app:rescaling}.

We will break up our analysis into two subsections. First, we study the parametric weak coupling limit, i.e. a running dilaton $\tau_I=e^{-\phi} \to \infty$, in the following subsection. After that we will consider the case where the dilaton is stabilized. We will try to reproduce settings from \cite{Calderon-Infante:2022nxb} that were argued to provide asymptotic accelerated expansion.

\subsection{Asymptotically weak coupling} \label{ssec:weakcouling}
We first show that it is impossible to get accelerated expansion when the dilaton is rolling in the scalar potential arising from \eqref{eqn:superpotential} and \eqref{eqn:Kahlerpotential} (see \cite{Shiu:2023nph} for a more generic argument). In terms of the real fields, $\tau = \tau_{R} + \rmi \tau_{I}$ and $U_{j} = U_{j R} + \rmi U_{j I}$, where $j=1,2,3$, the potential can be written as
\be
V = \frac{A_{\tau^2}}{\tau_{I}^2} - \frac{A_{\tau^3}}{\tau_{I}^3} + \frac{A_{\tau^4}}{\tau_{I}^4}, 
\ee
where  $A_{\tau^2}$, $A_{\tau^3}$ and $A_{\tau^4}$ are functions of the fluxes, the axions ($U_{1 R}, U_{2 R}, U_{3 R}, \tau_{R}$) and the complex structure moduli ($U_{1 I}, U_{2 I}, U_{3 I}$). Here in the limit $\tau_{I} \rightarrow \infty$, the $\frac{1}{\tau_{I}^2}$ term dominates and leads to $\gamma = \sqrt{2}$ if all fields are stabilized  and $\gamma > \sqrt{2}$ otherwise (see equation \eqref{eq:epsilonrhotau} above). If $A_{\tau^2}$ is zero or other fields roll also in such a way that the $A_{\tau^2}$ term does not dominate, then the potential would be even steeper and accelerated expansion certainly cannot happen.

If we are able to actually stabilize all the axions and the complex structure moduli such that the $A_{\tau^2}$ term dominates, then one can examine how the subleading term $-A_{\tau^3}/\tau_I^3$ affects the total number of e-folds we can get. This term is arising from localized sources and is negative for a net number of O3-planes. That means it would make the asymptotic potential for the dilaton from the $A_{\tau^2}/\tau_{I}^2$ flatter.

Unfortunately, one can show that one cannot stabilize the axions and all the complex structure moduli while keeping the potential positive. To see this we employ the rescaling discussed earlier and set $U_{1}=U_{2}=U_{3}=\rmi$ at the putative minimum. Since $\tau_{I}$ is a rolling field, we rescale the fluxes so that $\tau_{R}=0$ only and do not rescale $\tau_I$. The potential value at the minimum is then given by 
\ba\label{eq:Vmin}
   V &= \frac{\sum_{a=0}^3((h_a)^2+(h^a)^2)+6(h^0 (h_1+h_2+h_3)-h_0 (h^1+h^2+h^3)-h_1 h_2 -h_1 h_3-h_2 h_3-h^1h^2-h^1 h^3-h^2 h^3)}{128 \tau_{I}^{2}} \nonumber\\[2mm]
   &\quad - \frac{\sum_{a=0}^3(f^a h_a - f_a h^a)}{16 \tau_{I}^{3}}+ \frac{\sum_{a=0}^3 ((f_a)^2 + (f^a)^2)}{32 \tau_{I}^{4}}\,.
\ea
The first derivatives with respect to $U_{jI}, U_{jR}, \tau_{R}$ for $j=1,2,3$ must vanish. Since we are interested in the limit $\tau_I \to \infty$ we can simplify the equations by expanding them around large $\tau_I$, keeping the first non-trivial order in $\tau_I$. We find
\begin{align}
    \partial_{\tau_R} V &= -\frac{\sum_{a=0}^3(f_a h_a+f^a h^a)}{16 \tau_I^4}\,,\cr
    \partial_{U_{jR}} V &= \frac{3 \sum_{a=0}^3 h_a h^a -6 h^j h_j + h_0 h_j-h^j h^0-h_k h^l-h_l h^k}{64 \tau_I^2}+ \mathcal{O}(\tau_I^{-4})\,,\\
    \partial_{U_{jI}} V &=\frac{\sum_{a=0}^3((h^a)^2\!-(h_a)^2)\!-2(h^j)^2\!+2(h_j)^2\!+6 (h_0 h^j+h_j h^0+h_k h_l-h^k h^l)}{128 \tau_I^2}+ \mathcal{O}(\tau_I^{-4}) \,,\nonumber
\end{align}
where $k,l \in \{1,2,3\}, k\neq l \neq j \neq k$.

The derivative $\partial_{\tau_R} V=0$ can be solved via the $f_a$ or $f^a$ fluxes that do not appear at leading order in the scalar potential for large $\tau_I$. Thus, we can neglect this equation. We solve the three equations $\partial_{U_{jR}}V=0$ for $j=1,2,3$ in terms of the three fluxes $h_j$, finding a unique solution. We then plug this solution into $\partial_{U_{1I}} V=0$ and solve it in terms of $h^0$. There are two solutions for $h^0$. Both of these solutions actually ensure that $\partial_{U_{2I}} V=\partial_{U_{3I}} V=0$ as well, so that we have no further equations to solve. 

For both of the solutions, we find that the fluxes fixed by the solutions are real, iff
\begin{equation}\label{eq:constraint}
    (-7 h_0 + h^1 + h^2 + h^3)(h_0 - 7 h^1 + h^2 + h^3) (h_0 + 
    h^1 - 7 h^2 + h^3) (h_0 + h^1 + h^2 - 7 h^3) < 0\,.
\end{equation}
We then calculate the value of $V$ at the minimum and find the same flux-dependent expression for both solutions. If the above constraint in equation \eqref{eq:constraint} is imposed then it follows that $V_{min}<0$. So, the putative minimum after stabilizing the complex structure moduli is AdS. There are a few special cases to further be considered if the fluxes take on special values or vanish. We discuss these in the appendix \ref{app:special} and find that these are not viable either. Thus, there is no solution in this model that has a rolling dilaton with $\tau_I=e^{-\phi} \to \infty$ and $\gamma=\sqrt{2}$.

If both the dilaton and some or all of the complex structure moduli are to roll, then we cannot find $\gamma \leq \sqrt{2}$ as any direction that the fields would roll along would have to be as steep or steeper than the $\tau_{I}$ direction. The reason is that $\tau_I$ appears in the denominators of \emph{all} terms of the scalar potential and in all of them the corresponding $\gamma$ for $\tau_I$ is greater or equal to $\sqrt{2}$. So in order to ever get $\gamma \leq \sqrt{2}$ we need to stabilize the dilaton and investigate the scenario where the complex structure moduli roll, which is what we are doing in the next subsection.

\subsection{Finite coupling}
The dependence of the scalar potential on the complex structure moduli is shown below 
\ba
    V &=& A_{1} U_{1I} U_{2I} U_{3I} + A_{2} \frac{U_{1I}U_{2I}}{U_{3I}} + A_{3} U_{1I} +A_{4} \frac{U_{1I} U_{3I}}{U_{2I}} +A_{5} U_{2I} + A_{6} U_{3I}  \nonumber \\[2mm]
    && + A_{7} +A_{8} \frac{U_{1I}}{U_{2I} U_{3I}} +A_{9} \frac{U_{2I} U_{3I}}{U_{1I}}  +A_{10} \frac{U_{2I}}{U_{1I}U_{3I}}+A_{11} \frac{U_{3I}}{U_{1I} U_{2I}} + A_{12} \frac{1}{U_{1I} U_{2I} U_{3I}} \nonumber\\[2mm]
    && + A_{13} \frac{1}{U_{1I}} + A_{14} \frac{1}{U_{2I}} +A_{15} \frac{1}{U_{3I}}\,, 
\ea
where the $A_{i}$'s are functions of the dilaton, the axions, and the fluxes. We will stabilize the axio-dilaton and due to the flux rescaling discussed above we assume that the minimum is at $\tau_{R}=0, \tau_{I}=1$. Note that this does not mean that the solution is at strong coupling. Neglecting flux quantization any solution can be shifted to this point and one can always go to weak coupling by a flux rescaling (modulo constraints coming from flux quantization and the tadpole cancellation condition). Similarly, we can set $U_{1R}=U_{2R}=U_{3R}=0$ by shifting the fluxes. 

There is a permutation symmetry between the three complex structure moduli in the above scalar potential. Given this, we can assume without loss of generality the following rates of growth of the fields $U_{1I} \geq U_{2I} \geq U_{3I}$ in any asymptotic limit. Such asymptotic limits could have either $U_{1I}$ going to infinity or $U_{1I}, U_{2I}$ going to infinity or all three $U_{jI}$'s going to infinity. We see then that $A_{1}, A_{2}, A_{3}$ have to be zero as these terms necessarily blow up asymptotically. This can be achieved by setting $f^{0}=h^{0}=f^{3}=h^{3}=0$ in $W$ in equation \eqref{eqn:superpotential}. This in turn actually implies that $A_5=0$ as well and the scalar potential reduces to
\ba\label{eq:Potentialfinite}
    V &=& A_{4} \frac{U_{1I} U_{3I}}{U_{2I}} + A_{6} U_{3I} + A_{7} +A_{8} \frac{U_{1I}}{U_{2I} U_{3I}} +A_{9} \frac{U_{2I} U_{3I}}{U_{1I}}  +A_{10} \frac{U_{2I}}{U_{1I}U_{3I}} \nonumber \\[2mm]
    && +A_{11} \frac{U_{3I}}{U_{1I} U_{2I}} + A_{12} \frac{1}{U_{1I} U_{2I} U_{3I}} + A_{13} \frac{1}{U_{1I}} + A_{14} \frac{1}{U_{2I}} +A_{15} \frac{1}{U_{3I}}\,. 
\ea
The term $A_7$ is the contribution from localized sources like O3-planes and D3-branes and can be written in terms of the flux parameters using the tadpole cancellation condition in equation \eqref{eq:tadpole} above. It is proportional to 
\begin{equation} \label{eq:A7}
    A_7 \propto \frac{2 N_{D3}-N_{O3}}{\tau_I^3}\,.
\end{equation}
One might be tempted to set it to zero to avoid having to deal with its constant asymptotic value. However, it is the only term in the potential that scales like $\tau_I^{-3}$ and it is crucial for stabilizing the dilaton, which can be seen as follows: 
The scalar potential with contributions from the NSNS-sector, localized sources, and the RR-sector take the schematic form as a function of $\tau_I$ (cf. equation \eqref{eq:Vmin} above)
\be \label{eqn:schematicpotential}
   V = \frac{A}{\tau_I^{2}} + \frac{B}{\tau_I^{3}} + \frac{C}{\tau_I^{4}}\,.
\ee
Let us assume $B \propto A_7 = 0$ and assume we have rescaled $A$ and $C$ so that the extremum is at $\tau_I=1$. Then $\partial_{\tau_I} V=0$ implies that $A=-2C$. At this extremum, the potential and its second derivative have opposite signs
\ba
   V |_{A=-2 C} &=& - C\,,\nonumber\\[2mm]
   \frac{d^{2} V}{d\tau_I^{2}} |_{A=-2 C} &=& 8 C\,.
\ea
So, regardless of whether we can find an interesting asymptotic trajectory in the potential in \eqref{eq:Potentialfinite} or not, we cannot stabilize the dilaton and keep the potential positive asymptotically if $A_7=0$. So, we have to keep $A_7$ in the potential. 

Having established the importance of the $A_7$ term, we now explore two different options of canceling it asymptotically to ensure that we approach a Minkowski solution. Let us first assume that all three $U_{jI}$ are rolling to infinity. This means that in the scalar potential in equation \eqref{eq:Potentialfinite} we have to also set $A_4=A_6=0$, which can be achieved by setting $f^{2}=h^{2}=0$. This leaves us with 
\ba\label{eq:Potentialfinite2}
    V &=& A_{7} +A_{8} \frac{U_{1I}}{U_{2I} U_{3I}} +A_{9} \frac{U_{2I} U_{3I}}{U_{1I}}  +A_{10} \frac{U_{2I}}{U_{1I}U_{3I}} +A_{11} \frac{U_{3I}}{U_{1I} U_{2I}} + A_{12} \frac{1}{U_{1I} U_{2I} U_{3I}} \nonumber \\[2mm]
    &&  + A_{13} \frac{1}{U_{1I}} + A_{14} \frac{1}{U_{2I}} +A_{15} \frac{1}{U_{3I}}\,. 
\ea
Given our hierarchy $U_{1I} \geq U_{2I} \geq U_{3I}$ we see from the scalar potential above that only the terms $A_8$ and $A_9$ could be asymptotically non-zero. If and only if $U_{1I}$ scales like $U_{2I} U_{3I}$, do they both approach a constant that can cancel $A_7$. Let us assume the following generic parameterization for $\sigma \to \infty$: $U_{1 I} = \sigma$, $U_{2 I} = \alpha_2 \sigma^{\beta}$ and $U_{3 I} = \alpha_3 \sigma^{1-\beta}$, with $\tfrac12 \leq \beta < 1$, $\alpha_2,\alpha_3>0$. Let us now calculate the derivatives of $V$ in equation \eqref{eq:Potentialfinite} with respect to the complex structure moduli in this limit to leading order
\begin{align}\label{eq:firstderivatives}
    \partial_{U_{1 I}} V &= A_{8} \frac{1}{U_{2I} U_{3I}} -A_{9} \frac{U_{2I} U_{3I}}{U_{1I}^2}  +\ldots = \lp \frac{A_{8}}{\alpha_2 \alpha_3} - A_{9} \alpha_2 \alpha_3\rp \sigma^{-1} +\ldots \nonumber \\[2mm]
    \partial_{U_{2I}} V &= -A_{8} \frac{U_{1I}}{U_{2I}^2 U_{3I}} +A_{9} \frac{U_{3I}}{U_{1I}}+\ldots = -\frac{1}{\alpha_2} \lp \frac{A_{8}}{\alpha_2 \alpha_3} - A_{9} \alpha_2 \alpha_3 \rp \sigma^{-\beta} +\ldots \\[2mm]
    \partial_{U_{3I}} V &= -A_{8} \frac{U_{1I}}{U_{2I} U_{3I}^2} +A_{9} \frac{U_{2I}}{U_{1I}}+\ldots = -\frac{1}{\alpha_3} \lp \frac{A_{8}}{\alpha_2 \alpha_3^2} - A_{9} \alpha_2 \rp \sigma^{\beta-1} +\ldots \nonumber
\end{align}
Given the simple diagonal field space metric arising from the \K potential in equation \eqref{eqn:Kahlerpotential} above, we have to demand that all derivatives are negative in order to ensure that the three scalar fields actually do roll to infinity and not away from it. Since $\alpha_2>0$ and $\alpha_3>0$ we clearly have found a contradiction. It is not possible for all derivatives to be negative. This means that all three complex structure moduli cannot roll asymptotically to infinity and cancel the contribution from the localized sources, i.e. the $A_7$ term.

As a result, we now focus on the scenario where two of the complex structure moduli are rolling and the third one $U_{3I}$ along with the dilaton and the axions needs to be stabilized. This allows us to use the terms $A_{6} U_{3I}$ and $A_{15}/U_{3I}$ in the scalar potential in equation \eqref{eq:Potentialfinite} to cancel the $A_7$ term to find asymptotic Minkowski solutions. I.e. at the minimum for $\tau$ and $U_{3}$ we would have $A_{6} U_{3I}+A_7+A_{15}/U_{3I}=0$. Assuming this, we can rewrite the scalar potential above in equation \eqref{eq:Potentialfinite} as follows
\be \label{potential2fields}
    V = \hat{A}_{1} \frac{U_{1 I}}{U_{2 I}} + \hat{A}_{2} \frac{U_{2 I}}{U_{1 I}} + \hat{A}_{3} \frac{1}{U_{1 I}} + \hat{A}_{4} \frac{1}{U_{2 I}} + \hat{A}_{5} \frac{1}{U_{1 I} U_{2 I}}\,,
\ee
where $\hat{A}_i$ are now functions of the axions, $\tau_I$ and $U_{3I}$. As before, without loss of generality, we assume the rates of growth are ordered, $U_{1 I} \geq U_{2 I}$. For the following parameterization $U_{1 I} = \alpha \sigma^{\beta}, U_{2 I} = \sigma$ with $\beta \geq 1$ we get from equation \eqref{eqn:slowroll} above that
\begin{align}
    \alpha \beta \sigma^{\beta-1} \frac{d\sigma}{dt} &= \frac{2}{\sqrt{3V}} (\hat{A}_{1} \alpha^2   \sigma^{2 \beta-1} - \hat{A}_{2} \sigma - \hat{A}_{3} - \hat{A}_{5} \sigma^{-1}) \,, \\[2mm]
    \frac{d\sigma}{dt} &= \frac{2}{\sqrt{3V}} (-\hat{A}_{1} \alpha \sigma^{\beta} + \hat{A}_{2} \alpha^{-1} \sigma^{2-\beta}-\hat{A}_{4} - \hat{A}_{5} \alpha^{-1} \sigma^{-\beta})\,.\nonumber
\end{align}
Taking the ratio of these two equations we find
\be
    \beta = \frac{\hat{A}_{1} \alpha \sigma^{\beta} - \hat{A}_{2} \alpha^{-1} \sigma^{2-\beta}-\hat{A}_{3} \alpha^{-1} \sigma^{1-\beta}-\hat{A}_{5} \alpha^{-1} \sigma^{-\beta}}{-\hat{A}_{1} \alpha \sigma^{\beta} + \hat{A}_{2} \alpha^{-1} \sigma^{2-\beta}-\hat{A}_{4} - \hat{A}_{5} \alpha^{-1} \sigma^{-\beta}}\,.
\ee
For $\sigma \rightarrow \infty$, we examine two cases. For $\beta>1$, the right-hand side  evaluates to $-1$ which is a contradiction. And for $\beta = 1$, again the right-hand side evaluates to be $-1$. So, we return to \eqref{potential2fields} and set $\hat{A}_{1} = 0$. This requires us to set the following fluxes in equation \eqref{eqn:superpotential} for $W$ to zero $f^2=h^2=f_1=h_1=0$. Following the same procedure as before we get
\be
\beta = \frac{-\hat{A}_{2} \alpha^{-1} \sigma^{2-\beta}- \hat{A}_{3} \alpha^{-1} \sigma^{1-\beta}- \hat{A}_{5} \alpha^{-1} \sigma^{-\beta}}{\hat{A}_{2} \alpha^{-1} \sigma^{2-\beta} - \hat{A}_{4} - \hat{A}_{5} \alpha^{-1} \sigma^{-\beta}}\,.
\ee
For $1 \leq \beta<2$, in the limit $\sigma \rightarrow \infty$, the right-hand side evaluates to $-1$ which is a contradiction to our initial assumption. For $\beta>2$, the limit of the right-hand side evaluates to $0$ which is again a contradiction. For $\beta=2$ we get, 
\be
    2 = \frac{-\hat{A}_{2} \alpha^{-1}}{\hat{A}_{2} \alpha^{-1} - \hat{A}_{4}}\,,    
\ee
which gives us $\alpha = \frac{3 \hat{A}_{2}}{2 \hat{A}_{4}}$. This in turn uniquely fixes the trajectory to, $U_{1 I} = \frac{3 \hat{A}_{2}}{2 \hat{A}_{4}} \sigma^{2}, U_{2 I} = \sigma$. For this trajectory we find $\gamma = \sqrt{\frac{2}{5}} < \sqrt{2}$. However, setting $f^2=h^2=f_1=h_1=0$ sets $\hat{A}_{4}=0$ and hence this trajectory is not realisable. 

There are no further potential trajectories with two fields rolling to infinity. Similarly, one cannot get accelerated expansion from a single monomial with a single field rolling to infinity. Thus, we have proven that the loophole pointed out in \cite{Calderon-Infante:2022nxb} cannot be realized in these simple non-geometric models with four moduli. We will now proceed and discuss some asymptotic behaviors for the scalar potentials that can arise in asymptotic limits.

\subsection{Asymptotic ridges and valleys}\label{ssec:ridges}
For exponential scalar potentials, it is natural to have a generalized weak coupling region in which the scalar potential goes to zero if one sends a scalar field, not necessarily the dilaton, to infinity. In a corresponding generalized strong coupling region, the scalar potential grows exponentially. 
Above we have been studying generalized weak coupling limits in which we send combinations of scalar fields to infinity, while the scalar potential approaches zero. 
However, in some cases we encountered terms that did not vanish in our particular asymptotic limit, leaving finite values for the scalar potential. One might ask the more general question 
of the possible behaviors of scalar potentials in asymptotic limits. Clearly in principle everything ranging from $-\infty$ to $+\infty$ could be a possible asymptotic value. However, zero is being singled out as the value for which we can trust the scalar potential even in the asymptotic limit. The reason is the Distance Conjecture \cite{Ooguri:2006in}, which states that for any infinite distance limit in scalar moduli space, $\phi \to \infty$, there is a tower of states which become light, with masses given by $m(\phi) \sim e^{-\alpha \phi}$. This leads to a decreasing species scale, i.e. a decreasing UV cutoff of the low energy effective theory \cite{Hebecker:2018vxz} (see also \cite{Hamada:2021yxy, vandeHeisteeg:2023ubh, Andriot:2023isc, vandeHeisteeg:2023uxj} for recent related work). Thus, in the asymptotic limit any scalar potential that does not go to zero will describe an EFT that necessarily breaks down after a finite distance.\footnote{While this should be obvious for $V>0$, since the potential energy cannot be larger than the cutoff of the theory, it should also apply to negative $V<0$. Indeed, the energy scale associated with the scalar potential should have a magnitude below the cutoff in order for the EFT to make sense in AdS.}

Assuming that the Distance Conjecture is correct, finite values for the scalar potential in asymptotic limits are not trustworthy since the low energy effective theory is breaking down. However, one might still be able to study such a limit if
the tower of states that are becoming light can be integrated in. Even if that is not possible, it is also interesting to have a finite range of scalar field space in which the scalar potential approaches a constant value and therefore does not change much. For example, for models of inflation we need a sufficiently flat scalar potential, and one that exponentially approaches a constant could certainly be a viable candidate (see \cite{Dias:2018pgj} for related work). 
Thus, in this section we study asymptotic limits with finite scalar potential values for our scalar potential arising from $W$ and $K$ in equations \eqref{eqn:superpotential} and \eqref{eqn:Kahlerpotential}. Below we present two exemplary cases, one where $V$ approaches a finite positive value and one where it approaches a finite negative value. While these two examples arise in our particular non-geometric model, we obtained similar features in other geometric flux compactifications as well. Thus, we believe that these asymptotic limits are generic and not specific to non-geometric models.

Consider the following choice of fluxes 
\begin{align}
    &h^{0} = h^{1} = h^{2} = h^{3} = f^{0} = f^{2} = f^{3} = 0\,,\cr 
    &h_{1} = \frac{12}{f^{1}}\,,\quad f_{2} = \frac{f_{1} f^{1} h_{2}}{12}\,, \quad f_{3} = \frac{f_{1} f^{1} h_{3}}{12}\,.
\end{align} 
The tadpole cancellation condition is satisfied for the above choice without needing to add any D3-branes, i.e. $N_{D3}=0$ in equation \eqref{eq:tadpole}. 

We can solve the axionic first derivative equations $\partial_{\tau_R} V = \partial_{U_{j R}}=0$ and find that they are solved for
\begin{equation}
    \tau_{R} = \frac{f_{1} f^{1}}{12}\,, \qquad U_{1R} = -\frac{f^{1} h_{0}}{12}\,, \qquad U_{2R}=U_{3R}=0\,.
\end{equation}
Now let us look at an asymptotic limit, $\sigma \to \infty$, with
\begin{equation}
    U_{1I} = \sigma^2\,, \qquad U_{2I} = U_{3I} = \sigma   \,. 
\end{equation} 
In the large $\sigma$ limit there are two solutions when stabilizing the dilaton, $\partial_{\tau_I}V=0$, namely
\begin{equation}
    \tau_{I} = \frac{(f^{1})^2 \left(3\pm\sqrt{7}\right)}{6} + \mathcal{O}(\sigma^{-1})\,.
\end{equation}
The two solutions correspond to two different asymptotic constant values for the scalar potential,
\begin{align}
    V_{min} &= \frac{81 \left(2 \sqrt{7}+5\right)}{2 \left(\sqrt{7}+3\right)^4 (f^1)^6}+ \mathcal{O}(\sigma^{-1}) \qquad \text{for}\qquad \tau_{I} = \frac{(f^{1})^2 \left(3+\sqrt{7}\right)}{6} + \mathcal{O}(\sigma^{-1})\,, \\
    V_{min} &= -\frac{81 \left(2 \sqrt{7}-5\right)}{2 \left(3-\sqrt{7}\right)^4 (f^1)^6}+ \mathcal{O}(\sigma^{-1}) \qquad \text{for}\qquad \tau_{I} = \frac{(f^{1})^2 \left(3-\sqrt{7}\right)}{6} + \mathcal{O}(\sigma^{-1})\,.\nonumber
\end{align}
Thus, this single scalar potential has two asymptotic limits in which the dilaton is extremized. In one of them, the scalar potential takes on a positive constant value, and in the other a negative one.

\begin{figure}[htbp]
\centerline{\includegraphics[width=.757\textwidth]{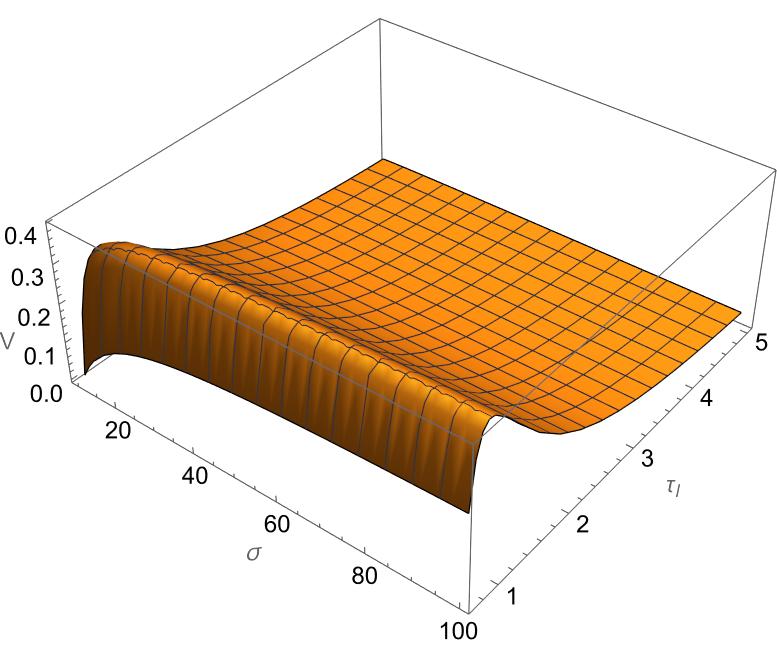}}
\caption{Asymptotically for large $\sigma$ a ridge emerges in the scalar potential.}
\label{fig:ridge}
\end{figure}

In figure \ref{fig:ridge} we show the ridge-like shape that appears for positive values of the scalar potential. For the plot, we have set all the free fluxes equal to 1, i.e. $f^1 = f_0 = f_1 = h_0 = h_2 = h_3 =1$.
The figure shows that the scalar potential approaches its asymptotic positive value $V(\sigma \to \infty, \tau = (3+\sqrt{7})/6) \approx .41$ from below. This means that the complex structure moduli $U_{1I}=\sigma^2$, $U_{2I}=U_{3I}=\sigma$ would actually not roll to $\sigma=\infty$ along this ridge but rather to small $\sigma$  values. Furthermore, we see that the dilaton direction $\tau_I$ is a maximum. The $\eta_V$ value for $\tau_I$ on this ridge is larger than one and one cannot use this ridge to obtain slow-roll inflation. Nevertheless, it is interesting -- and probably known --  that such ridges can arise in asymptotic limits of the scalar field space. For large values of $\sigma$ one expects that the species scale or the masses of some tower of states drop below the value of the scalar potential and the effective low energy theory breaks down.

\begin{figure}[htbp]
\centerline{\includegraphics[width=.75\textwidth]{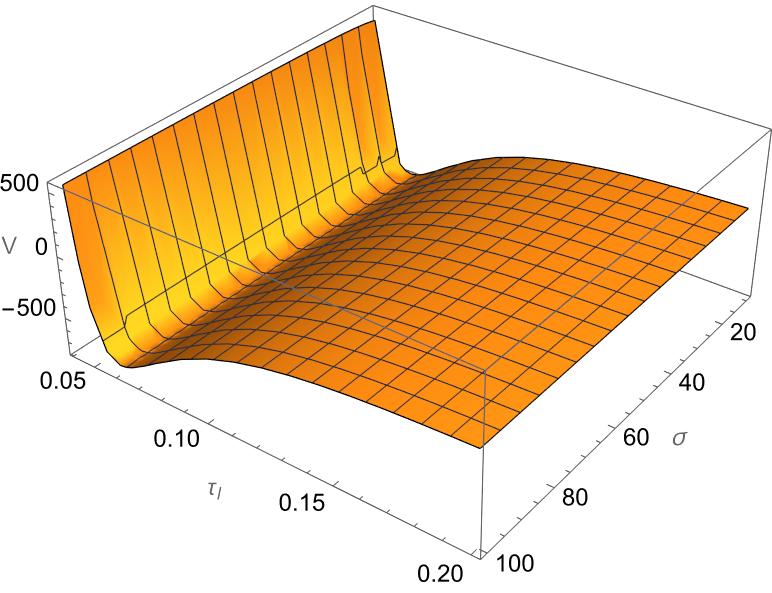}}
\caption{Asymptotically for large $\sigma$ a valley emerges in the scalar potential at smaller values of the string coupling $\tau_I$.}
\label{fig:valley}
\end{figure}

Let us now follow the scalar potential above to smaller $\tau_I$ values, where we encounter a valley-like shape that is shown in figure \ref{fig:valley}.
We see that the scalar potential asymptotes in the valley to a constant negative value for large $\sigma$. This time the dilaton $\tau_I$ is stabilized at a minimum. Again, as before the complex structure moduli $U_{1I}=\sigma^2$, $U_{2I}=U_{3I}=\sigma$ will not asymptotically roll to large $\sigma$ but rather away from it since the potential approaches its asymptotic value from below. However, contrary to the positive asymptotic ridge, here all axions and the dilaton have positive masses and are stabilized in a minimum. 

This negative AdS valley seems at first sight in tension with the recently proposed Anti-Trans-Planckian Censorship Conjecture (ATCC) \cite{Andriot:2022brg}, which states that negative scalar potentials  with positive slope should also go to zero exponentially in any asymptotic limit. However, as in the case above with a positive asymptotic value for the scalar potential, we expect here as well that the low energy effective theory breaks down at some finite value for $\sigma$, when the species scale or the masses of light towers become of the same order as the absolute value of the scalar potential. Thus, it seems plausible that this breakdown of the EFT prevents a violation of the ATCC. This is an interesting point that deserves further study in the future.

After the first version of this paper appeared, the authors of  \cite{Andriot:2025gyr} informed us that it is not accurate to characterize the limits in this subsection as ridges and valleys. In the full scalar field space spanned by $\tau, U_{1I}, U_{2I}, U_{3I}$ there is a direction along which the gradient is non-vanishing even in the asymptotic limit $\sigma \to \infty$. Thus, these limits are actually slopes rather than ridges or valleys.

\section{Relating the Distance and Strong Asymptotic dS Conjecture}\label{sec:DistanceStrongdS}
It was discussed previously  in \cite{Etheredge:2022opl} that $\gamma_{min}= 2/\sqrt{d-2}$, the minimal coefficient allowed  by the Strong Asymptotic dS Conjecture, is twice the minimal value allowed by the Sharpened Distance Conjecture \cite{Ooguri:2006in, Klaewer:2016kiy, Etheredge:2022opl}. The (refined) Distance Conjecture \cite{Ooguri:2006in, Klaewer:2016kiy} states that for any infinite distance limit in scalar moduli space, $\phi \to \infty$, there is a tower of states becoming light, with $m(\phi) \sim e^{-\alpha \phi}$. The Sharpened Distance Conjecture \cite{Etheredge:2022opl} furthermore states that $\alpha \geq \alpha_{min} = 1/\sqrt{d-2}$. The intriguing relationship $\gamma_{min}=2\alpha_{min}$ was also discussed previously  in \cite{Hebecker:2018vxz, Andriot:2020lea, Bedroya:2020rmd}. It was shown in \cite{Etheredge:2022opl} that the Sharpened Distance Conjecture together with the Emergent String Conjecture \cite{Lee:2019wij, Lee:2019xtm} implied the Strong Asymptotic dS Conjecture. Here we give an intuitive argument for the opposite direction. If this argument is correct and can be generalized, then the Strong Asymptotic dS Conjecture would imply the Sharpened Distance Conjecture with the precise relationship $\gamma_{min}=2\alpha_{min}$ above.

Let us quickly recall that for a free 5d scalar $\Phi(x^\mu,y)$ on $\mathbb{R}^{3,1} \times S^1$ we can write the action in 5d or in 4d via
\begin{align}\label{eq:freescalar}
    S &= -\frac12 \int d^4x \, dy \sqrt{-g_5}\, \partial_m \Phi \partial^m \Phi \cr
    &= -\frac12 (2\pi R)\int d^4x \sqrt{-g_4} \, \sum_{n\geq0} \left(\partial_\mu  \Phi_n \partial^\mu \Phi_n + \left(2\pi n\right)^2 \Phi_n^2\right)\cr
    &= -\frac12 \int d^4x \sqrt{-g_4^{E}} \,\sum_{n\geq0} \left(\partial_\mu  \Phi_n \partial^\mu \Phi_n + \left(\frac{n}{R}\right)^2 \Phi_n^2\right)\,.
\end{align}
In compactifications to $d$ dimensions, one usually drops the heavy state $\Phi_n$ for $n\neq 0$ but this is not necessary or required.\footnote{String theory provides us with a UV complete theory. In simple toroidal orbifolds, one can calculate the full mass spectrum and keep track of all fields. There is no reason to discard for example KK-modes or higher string excitations. In more complicated compactifications we cannot calculate the complete mass spectrum but our argument below is independent of the explicit knowledge of all the massive states.} Having a lower dimensional $d$-dimensional theory one could have likewise kept track of heavier states $\Phi_n$ like the KK-towers that arise in the compactification and the massive string excitations. This would then modify the $d$-dimensional scalar potential $V_d$ by additional new terms as follows,\footnote{It is an interesting question at which energy scale the low energy EFT breaks down, see for example \cite{Montero:2022prj, Rudelius:2022gbz}. Here we include all the fields in all the towers, sending the UV cutoff to infinity.}
\begin{equation}
    V = V_d(\phi) + \sum_{\rm t} \left(\sum_{n\geq0} \left(m^{(t)}_n(\phi)\right)^2 (\Phi^{(t)}_{n})^2 \right)\,,
\end{equation}
where $t$ labels the different towers. Recall that the logic of \cite{Ooguri:2006in} states that for any infinite distance limit $\phi \to \infty$ there is at least one tower that satisfies $m(\phi) \sim e^{-\alpha \phi}$.\footnote{It does not matter for our argument whether this direction corresponds to a massless modulus, i.e. whether $V_d(\phi)=0$ or not. If $V_d(\phi)\neq0$, we are dealing with the refined Distance Conjecture \cite{Klaewer:2016kiy}.} Keeping only track of towers $\tilde t$ that become light for $\phi \to \infty$ we have 
\begin{equation}
    V = V_d(\phi) + \sum_{\rm \tilde{t}} \left(\sum_{n\geq0} \left(m^{(\tilde t)}_n\right)^2 e^{-2\alpha^{(\tilde t)} \phi} (\Phi^{(\tilde t)}_{n})^2 \right)\,.
\end{equation}
Now the Sharpened Distance Conjecture states that the \emph{lightest} tower in any infinite distance limit $\phi \to \infty$ in $d$ dimensions satisfies $\alpha_{min} \geq \frac{1}{\sqrt{d-2}}$ \cite{Etheredge:2022opl}. In the above scalar potential the masses appear quadratically, leading naturally to exponentially decaying terms 
$ \sim e^{-\gamma^{(\tilde t)} \phi}$
for $\phi \to \infty$, with $\gamma^{(\tilde t)} =2\alpha^{(\tilde t)}$. 
From the Strong Asymptotic dS Conjecture we can then constrain $2\alpha^{(\tilde t)} \geq \gamma_{min} = 2/\sqrt{d-2}$. While the Distance Conjecture is very well established and tested, it is amusing to note that it actually could be implied by the absence of asymptotic accelerated expansion via the above argument.

The attentive reader will of course object to the above reasoning because of the $(\Phi^{(\tilde t)}_{n})^2$ factor in the above terms. Asymptotically in the above example $\Phi^{(\tilde t)}_{n}=0$ and the new terms in the scalar potential all vanish. Thus, the above argument does not go through that easily. However, when including massive fields from the tower we also need to include their interactions and once those are included it is not clear at all that $\Phi^{(\tilde t)}_{n}=0$ is the only possible solution for the $\Phi^{(\tilde t)}_{n}$.

Thus a generic expectation for the $d$-dimensional scalar potential in the asymptotic limit $\phi \to \infty$ would be 
\begin{equation}
    V = V_d(\phi) + \sum_{\rm \tilde{t}} \left(\sum_n \left(m^{(\tilde t)}_n\right)^2 e^{-2\alpha^{(\tilde t)} \phi}\, (\Phi^{(\tilde t)}_{n})^2 \right) + V_{int}(\phi, \Phi^{(\tilde t)}_n)\,.
\end{equation}
From the above scalar potential, it seems more reasonable to assume that the Strong Asymptotic dS Conjecture should apply also when including all massive towers and their interactions. 

Given that we are very much used to integrating out heavy states and setting the above $\Phi_n^{(t)}$ to zero, one might wonder how easy or hard it is to generate a non-trivial scalar potential which has a minimum for $\Phi_n^{(t)}$ away from zero. We will give here a simple toy example as a proof of principle that it is indeed possible to do this. Let us return to the above case of a real 5d scalar on $\mathbb{R}^{3,1} \times S^1$ and add to the action in equation \eqref{eq:freescalar} two exponential terms\footnote{Similar actions arise in 10-dimensional supergravity for the dilaton.}
\begin{equation}
    S = \int d^4x \, dy \lp -\frac12 \partial_m \Phi \partial^m \Phi - \Lambda_1 e^{-\gamma_1 \Phi} - \Lambda_2 e^{-\gamma_2 \Phi}\rp  \,.
\end{equation}
For a real scalar field we can make the following Ansatz
\begin{equation}
    \Phi(x^\mu,y) = \Phi_0(x^\mu)+ \sqrt{2} \sum_{n=1}^\infty \Phi_{n}(x^\mu) \cos(2\pi n y) = \Phi_0(x^\mu)+ \sqrt{2} \Phi_{1}(x^\mu) \cos(2\pi y) + \ldots
\end{equation}
For ease of presentation and because $\Phi_1$ is the lightest KK-mode, we set $\Phi_{n\geq2}=0$. We can then perform the integral over the $y$-direction and find
\begin{align}
    S &=\!\! \!\int \!\! d^4x \, dy \sqrt{-g_5} \bigg[\!\! -\!\frac12 \partial_m \!\left( \Phi_0(x^\mu)+ \sqrt{2} \Phi_{1}(x^\mu) \cos(2\pi y) \right) \! \partial^m \!\! \left(\Phi_0(x^\mu)+\! \sqrt{2} \Phi_{1}(x^\mu) \cos(2\pi y)\right) \cr
    &\qquad \qquad \qquad \quad- \Lambda_1 \, e^{-\gamma_1 (\Phi_0(x^\mu)+ \sqrt{2} \Phi_{1}(x^\mu) \cos(2\pi y))} - \Lambda_2 \, e^{-\gamma_2 (\Phi_0(x^\mu)+ \sqrt{2} \Phi_{1}(x^\mu) \cos(2\pi y))}\bigg] \cr
    &= \int d^4x \sqrt{-g_4^E}\; \bigg[ -\frac12 \left(\partial_\mu  \Phi_0 \partial^\mu \Phi_0 + \partial_\mu  \Phi_1 \partial^\mu \Phi_1 + \frac{1}{R^2} (\Phi_1)^2\right) \\
    &\qquad \qquad \qquad\quad - \frac{\Lambda_1}{4\pi^2 R^2} e^{-\gamma_1 \Phi_0} I_0(\sqrt{2} \gamma_1 \Phi_{1}) - \frac{\Lambda_2}{4 \pi^2 R^2} e^{-\gamma_2 \Phi_0} I_0(\sqrt{2} \gamma_2 \Phi_{1})\bigg) \bigg]\,.\nonumber
\end{align}
Thus, we see that interaction terms for the KK-modes appear naturally (here via the $I$-Bessel function). The Bessel function $I_0$ has the following expansion around $z=0$,
\begin{equation}
    I_0(z) = 1 +\frac{z^2}{4} +\frac{z^4}{64} + \frac{z^6}{2304} + \mathcal{O}(z^8)\,.
\end{equation}
We can thus easily choose the parameters $\gamma_i\geq 0$ and the $\Lambda_i$ with opposite signs to generate more minima for $\Phi_1$ at which $\Phi_1 \neq 0$. In a full model we would have to also stabilize $\Phi_0$ but that is the modulus that appears in the lower dimensional scalar potential anyways through a variety of terms. Note, that $1/R^2 = m(\phi)^2 \sim e^{-2\alpha \phi}$ actually multiplies \emph{all} terms in the scalar potential for $\Phi_1$. Thus, if we choose the parameters $\Lambda_i$ to generate a minimum with $\Phi_1 \neq 0$ then we can flow to $\phi \to \infty$ without disturbing this minimum.

While the above is only a toy model it supports the general idea that in a UV complete theory like string theory it is natural to include all massive fields and their interactions in the lower dimensional scalar potential. The absence of accelerated asymptotic expansion, as demanded by the Strong Asymptotic dS Conjecture, is then related to the Sharpened Distance Conjecture. Their prefactors naturally satisfy the same bound up to a factor of 2. The authors of \cite{Etheredge:2022opl} argued that the Sharpened Distance Conjecture combined with the Emergent String Conjecture \cite{Lee:2019wij, Lee:2019xtm} implies the Strong Asymptotic dS Conjecture. Together with our proposal this leads to an intriguing interconnection between these three swampland conjectures.

We should note, however, that it is not clear to us at the moment how to extend our argument to moduli spaces of more than one dimension.\footnote{We thank Tom Rudelius for pointing this out to us.} Since the Strong Asymptotic dS Conjecture only applies to gradient flows and the Sharpened Distance Conjecture to any direction in field space, one would need to ensure that the direction of interest is a gradient flow direction. Thus in the case where the massive tower of interest depends on several scalar fields one would have to be able to stabilize directions transverse to the flow in order for our argument to apply. We leave it to the future to check explicit examples and see whether this is possible or not.

\section{Conclusions}\label{sec:Conclusions}
Given the fact that our own universe is in a phase of accelerated expansion it is of paramount importance to understand how one can obtain similar cosmological histories from string theory. Given the difficulty of constructing explicit dS vacua that are widely accepted, a lot of work in the last several years has focused on the easier task of studying asymptotic limits of moduli space in string theory. In this paper we have extended previous work by studying asymptotic limits in a special non-geometric string compactification of type IIB string theory without \K moduli.

Our study was motivated by potential counter-examples to the Strong Asymptotic dS Conjecture \cite{Bedroya:2019snp, Rudelius:2022gbz} put forth in the recent paper \cite{Calderon-Infante:2022nxb}. The authors of that paper found counter-examples when neglecting the stabilization of the \K moduli and focusing only on the axio-dilaton and the complex structure moduli. Thus, our setup without \K moduli provides a natural way of realizing these counter-examples in full-fledged string theory constructions. However, it turns out that we cannot get sufficiently flat asymptotic regions in our scalar potential. It would be interesting to study this further in other non-geometric models and to extend the work of \cite{Calderon-Infante:2022nxb} to include \K moduli stabilization.

While we find no asymptotic regions in the scalar potential that could give rise to accelerated expansion, we do find limits in which the scalar potential takes on finite positive or negative values. In the case of a positive scalar potential the corresponding ridges\footnote{The characterization of these limits as ridges and valleys is misleading and they are rather slopes. See the discussion at the end of subsection \ref{ssec:ridges} and \cite{Andriot:2025gyr} for a discussion of similar limits.} in the scalar potential have large tachyonic directions. In the case of negative asymptotic values this is not the case. However, in both cases we find that these limits cannot be obtained dynamically since the scalar fields would want to roll away from these limits, towards the interior of moduli space.

Lastly, we discussed how the Strong Asymptotic dS Conjecture could potentially imply the Sharpened Distance Conjecture. We show that massive towers of states  appear naturally in the scalar potential and thus provide extra terms that are constrained by the Strong Asymptotic dS Conjecture. This leads naturally to a previously observed factor of 2 between the lower bounds on the exponential prefactors in these two conjectures. Our argument relies on the potential existence of non-trivial extrema for the massive fields in the towers of states that become light. While we provide a toy example that gives rise to such non-trivial extrema,  it would certainly be important to work out explicit examples to further check the connection we observed between the two conjectures. We hope to return to this in the near future.

\section*{Acknowledgments}
We like to thank David Andriot, Ben Heidenreich, Tom Rudelius, Flavio Tonioni and Irene Valenzuela for useful discussions. The work of S. C. and Y. T. is supported in part by the NSF grant PHY-2210271. The work of E. G., M. R. and T. W. is supported in part by the NSF grant PHY-2013988. M.R. acknowledges the support of the Dr. Hyo Sang Lee Graduate Fellowship from the College of Arts and Sciences at Lehigh University.
S.C. would like to thank the Harvard University Department of Physics and the Kavli Institute for Theoretical Physics for hospitality and support throughout part of this work.
This research was supported in part by the National Science Foundation under Grant No. NSF PHY-1748958.
T.W. would like to thank the LAPTh - CNRS for hospitality during the completion of this work. His visit was made possible thanks to the AAP USMB dSCordes (2023).

\appendix

\section{Rescaling the fluxes}\label{app:rescaling}
By giving up the quantization of fluxes temporarily and rescaling them, we can shift the axions to $0$ and the saxions to $1$ such that the superpotential in equation \eqref{eqn:superpotential} is invariant. Such a rescaling would lead to the rescaling of the scalar potential arising from equations \eqref{eqn:superpotential} and \eqref{eqn:Kahlerpotential}, by an overall positive factor. Consider a minimum of the scalar potential arising from \eqref{eqn:superpotential} and \eqref{eqn:Kahlerpotential} which is at $\tau = \tau_{0R}+\rmi \tau_{0I}\,,U_{j}=U_{0jR}+\rmi U_{0jI}$, where $j=1,2,3\,.$ We first shift the minimum to the point $\tau^\prime = \rmi \tau_{0I}\,, U_{j}^\prime=\rmi U_{0jI}$ via the following transformations of the fluxes,
\begin{align}
    &h^{0'} = h^{0}\,, \;\; 
 f^{0'}=f^{0}-h^{0} \tau_{0R}\,,\cr &h^{j'} = h^{j}-h^{0} \tau_{0R}\,, \cr
     &f^{j'} = f^{j} - U_{0jR} \tau_{0R} h^{0} - \tau_{0R} h^{j'} - U_{0jR} f^{0'}\,,\cr
    &h_{j}^{'} = h_{j} - \sum_{k\,,l}U_{0kR} h^{l}  + U_{0kR} U_{0lR} h^{0}\,,\cr
    &f_{j}^{'} = f_{j} - \tau_{0R} h_{j}^{'}  - \sum_{k\,,l}U_{0kR} f^{l}+U_{0kR} U_{0lR} f^{0} \,,\cr
    &h_{0}^{'} = h_{0} + \sum_{j=1}^{3}U_{0jR} h_{j}-\frac{1}{2}\sum_{j,k,l} \rho_{jkl} U_{0jR} U_{0kR} h^{l} + U_{01R} U_{02R} U_{03R} h^{0}\,,\cr
    &f_{0}^{'} = f_{0}  + \sum_{j}U_{0jR} f_{j}  -\frac{1}{2}\sum_{j,k,l} \rho_{jkl} U_{0jR} U_{0kR} f^{l} + U_{01R} U_{02R} U_{03R} f^{0} - \tau_{0R} h_{0}^{'}\,,
\end{align} 
where $j\,,k\,,l \in \{1,2,3\}, k\neq l \neq j \neq k$ and $\rho_{jkl} = 1$ iff $k\neq j \neq l \neq k$ and is vanishing otherwise. It can be easily checked that this leaves the superpotential in equation \eqref{eqn:superpotential} invariant,
\begin{align}
    W^\prime=& (f^{0'} - \tau' h^{0'}) U_{1}^{'} U_{2}^{'} U_{3}^{'} - (f^{1'}-\tau' h^{1'}) U_{2}^{'} U_{3}^{'}-(f^{2'}-\tau' h^{2'}) U_{1}^{'} U_{3}^{'}-(f^{3'}-\tau' h^{3'}) U_{1}^{'} U_{2}^{'} \nonumber\\[2mm]
   &+ (f_{1}^{'}-\tau' h_{1}^{'}) U_{1}^{'}+(f_{2}^{'}-\tau^{'} h_{2}^{'}) U_{2}^{'}+(f_{3}^{'}-\tau' h_{3}^{'}) U_{3}^{'} + (f_{0}^{'}- \tau' h_{0}^{'}) = W\,. 
\end{align}
Now to shift the minimum from $\tau^\prime = \rmi \tau_{0I}\,, U_{j}^\prime=\rmi U_{0jI}$ to $\hat{\tau} = \rmi\,, \hat{U_{j}}=\rmi$ for $j=1\,,2\,,3$ we require the following transformations,
\begin{align}
    &\hat{h}_{0} = h_{0}^{'} \tau_{0I} \,, \;\; \hat{f}_{0} = f_{0}^{'} \,,\cr
    &\hat{h}_{j} = h_{j}^{'}\tau_{0I} U_{0jI} \,, \;\;  \hat{f}_{j} = f_{j}^{'}U_{0jI} \,,\cr
    &\hat{h}^{j} = h^{j'}\tau_{0I} U_{0kI} U_{0lI} \,, \;\; 
 \hat{f}^{j} = f^{j'}U_{0kI} U_{0lI} \,,\cr
    &\hat{h}^{0} = h^{0'}\tau_{0I} U_{01I} U_{02I} U_{03I}\, ,   \;\; \hat{f}^{0} = f^{0'}U_{01I} U_{02I} U_{03I} \,,
\end{align}
where $j\,,k\,,l \in \{1,2,3\}, k\neq l \neq j \neq k$. Once again, it is easy to see that this leaves the superpotential in equation \eqref{eqn:superpotential} invariant
\begin{align}
    \hat{W}=& (\hat{f}^{0} - \hat{\tau} \hat{h}^{0}) \hat{U}_{1} \hat{U}_{2} \hat{U}_{3} - (\hat{f}^{1}-\hat{\tau} \hat{h}^{1}) \hat{U}_{2} \hat{U}_{3}-(\hat{f}^{2}-\hat{\tau} \hat{h}^{2}) \hat{U}_{1} \hat{U}_{3}-(\hat{f}^{3}-\hat{\tau} \hat{h}^{3}) \hat{U}_{1} \hat{U}_{2} \nonumber\\[2mm]
   &+ (\hat{f}_{1}-\hat{\tau} \hat{h}_{1}) \hat{U}_{1}+(\hat{f}_{2}-\hat{\tau} \hat{h}_{2}) \hat{U}_{2}+(\hat{f}_{3}-\hat{\tau} \hat{h}_{3}) \hat{U}_{3} + (\hat{f}_{0}- \hat{\tau} \hat{h}_{0}) = W\,. 
\end{align}

\section{Special cases at asymptotic weak coupling}\label{app:special}
Above in section 3.2 we described how a generic solution that extremizes the scalar potential with respect to the axions and complex structure moduli necessarily leads to an AdS minimum with $V_{min}<0$. However, there are a few special cases that need to be addressed to close all loopholes in our generic argument.

Above we solved the three equations $\partial_{U_{jR}}V=0$ for $j=1,2,3$ in terms of the three fluxes $h_j$. However, the $h_j$ appear in these equations with a certain prefactor,  and one needs to examine separately cases where, for example, one or all of these prefactors are zero. This leads to several special cases that we discuss below step by step.

The three axionic equations are of the form (see equations \eqref{eq:firstderivatives} above)
\begin{align}\label{eq:axderivatives}
0 = \partial_{U_{1R}} V &=\frac{1}{64 \tau_I^2} \lp h_1 (h_0-3 h^1) + \ldots \rp\,,\cr
0 = \partial_{U_{2R}} V &=\frac{1}{64 \tau_I^2} \lp h_2 (h_0-3 h^2) + \ldots \rp\,,\cr
0 = \partial_{U_{3R}} V &=\frac{1}{64 \tau_I^2} \lp h_3 (h_0-3 h^3) + \ldots \rp\,.
\end{align} 
If $h_0=3 h^1 =3 h^2=3 h^3$ we have a special case. Solving the equations $\partial_{U_{jR}} V =\partial_{U_{jI}} V = 0$ necessarily requires $h_0=0$. For $h_0=0$ we have $\partial_{U_{jR}} V = 0$ but we still need to solve $\partial_{U_{jI}} V = 0$. Up to permutations of the $j=1,2,3$ index there are four different solutions. For three of them the scalar potential at the minimum, $V_{min}$, is negative so we can discard them. The remaining case has
\be
h_0=h^1 =h^2=h^3 = 0\,,\quad h_1 =- h^0\,,\quad h_2 = h_3 = h^0\,.
\ee
The scalar potential at the minimum is given by
\be
V_{min} = \frac{(h^0)^2}{8 \tau_I^2} - \mathcal{O}(\tau_I^{-3}) \,.
\ee
This seems to be in principle interesting. We have a single leading term in the scalar potential that gives exactly $\gamma=\sqrt{2}$ from the dilaton $\tau_I=e^{-\phi}$ running to infinity (see equation \eqref{eq:epsilonrhotau}, which gives $\epsilon_V=1$). This is a setting right at the boundary and leads to a small number of e-folds. Now we however also have a subleading term that can further flatten the potential if it is negative (see equation \eqref{eq:A7} above). One could then hope to get a larger number of e-folds. However, if one calculates the Hessian of the scalar potential above one finds that in this case $U_{1I}$ is a tachyonic direction with $\eta_V=1$ along this tachyonic direction. Thus, there is no stable asymptotic trajectory in which only the dilaton $\tau_I=e^{-\phi}$ rolls.

Now let us assume without loss of generality that the first prefactor $(h_0-3h^1)$ in equation \eqref{eq:axderivatives} above is non-zero. We can then solve $\partial_{U_{1R}} V=0$ for $h_1$ and plug the result into $\partial_{U_{2R}} V=0$ and $\partial_{U_{3R}} V=0$ to get
\begin{align}
0 = \partial_{U_{2R}} V &=\frac{1}{64 \tau_I^2} \lp h_2 \frac{(h_0-h^3) (h_0-3 h^1-3 h^2+h^3)}{h_0-3 h^1} + \ldots \rp\,,\cr
0 = \partial_{U_{3R}} V &=\frac{1}{64 \tau_I^2} \lp h_3 \frac{(h_0-h^2) (h_0-3 h^1-3 h^3+h^2)}{h_0-3 h^1} + \ldots \rp\,.
\end{align} 
If the prefactors $(h_0-h^3) (h_0-3 h^1-3 h^2+h^3)$ and $(h_0-h^2) (h_0-3 h^1-3 h^3+h^2)$ both vanish then we have more special cases that we need to study in detail. We again find that solving all equations either leads to $V_{min}<0$ or a tachyonic direction with large $\eta_V$ value among the extremized complex structure moduli. 

If one of the prefactors is non-zero, we can assume without loss of generality that it is $(h_0-h^3) (h_0-3 h^1-3 h^2+h^3)$. Then we can solve $\partial_{U_{2R}} V=0$ for $h_2$ and plug the answer into $\partial_{U_{3R}} V=0$ to obtain
\begin{align}
0 = \partial_{U_{3R}} V =\frac{1}{64 \tau_I^2} &\lp \frac{h_3}{(h_0-h^3) (h_0-3 h^1-3 h^2+h^3)}\cdot C + \ldots \rp\,, \nonumber
\end{align} 
with
\begin{align}
C=& h_0^3 + 3 (h^1)^3 - 15 (h^1)^2 (h^2 + h^3)- 3 h_0^2 (h^1 + h^2 + h^3) \cr
&+ h^1 \lp -15 (h^2)^2 + 106 h^2 h^3 - 15 (h^3)^2 \rp + 3 (h^2 + h^3) \lp(h^2)^2 - 6 h^2 h^3 + (h^3)^2\rp\cr
&- h_0 \lp (h^1)^2 + (h^2)^2 - 6 h^2 h^3 + (h^3)^2 - 6 h^1 (h^2 + h^3)\rp\,.
\end{align} 
Again, if the prefactor $C$ of $h_3$ vanishes then we have another special case. This case is the hardest to analyze. In almost all subcases we find that $V_{min}<0$. There are a few cases where a positive scalar potential is possible. In some of those cases we could not prove in full generality that there is a tachyonic direction, because the resulting expressions are extremely complicated functions of the flux parameters. However, whenever we plugged in explicit values for the fluxes we always found that $V_{min}>0$ only arises in the presence of a tachyonic direction among the complex structure moduli. Thus, we have not found any interesting solutions with only the dilaton running to infinity in this special case either.

If $C\neq0$ we can solve $\partial_{U_{3R}} V =0$ for $h_3$. We then plug the three solutions for the $h_j$ into $\partial_{U_{1I}} V$ and find
\begin{equation}
0 = \partial_{U_{1I}} V =\frac{1}{128 \tau_I^2} \lp 4 (h^0)^2 \frac{A\cdot B}{C}  + \ldots \rp\,,
\end{equation}
with no linear term in $h^0$ and 
\begin{align}
    A&= \! (-7 h_0 + h^1 + h^2 + h^3)(h_0 - 7 h^1 + h^2 + h^3) (h_0 + 
    h^1 - 7 h^2 + h^3) (h_0 + h^1 + h^2 - 7 h^3)\,,\nonumber\\
    B&= h_0^2 - 6 h_0 h^1 + (h^1)^2 - (h^2)^2 + 6 h^2 h^3 - (h^3)^2\,.
\end{align}
Note that $A$ is the expression that appeared above in equation \eqref{eq:constraint}. If $A\cdot B \neq 0$, then we have the generic case discussed above in subsection \ref{ssec:weakcouling}. So, let us address the special cases $A=0$ and/or $B=0$.

We start out with the case where $A=0$. Up to permutations of $j=1,2,3$ there are two possibilities
\be
7 h_0= h^1 + h^2 + h^3\,, \qquad \text{or} \qquad 7 h_1= h_0 + h^2 + h^3\,.
\ee
Solving all the remaining equations in terms of the fluxes leads to a variety of solutions, all of which have $V_{min}<0$ and can therefore be discarded.

Lastly, we study the case $B=0$. Again there are many different subcases, most of which are leading to a negative scalar potential at the minimum. The ones that do lead to $V_{min}>0$ all have a tachyonic direction along one of the complex structure moduli $U_{jI}$ and thus do not allow for stable trajectories with only $\tau_I$ rolling. Thus, we have checked all special cases and excluded them.

\bibliographystyle{JHEP}
\bibliography{cite.bib}

\providecommand{\href}[2]{#2}\begingroup\raggedright\begin{thebibliography}{10}

\bibitem{SupernovaSearchTeam:1998fmf}
{\scshape Supernova Search Team} collaboration, \emph{{Observational evidence
  from supernovae for an accelerating universe and a cosmological constant}},
  \href{https://doi.org/10.1086/300499}{\emph{Astron. J.} {\bfseries 116}
  (1998) 1009} [\href{https://arxiv.org/abs/astro-ph/9805201}{{\ttfamily
  astro-ph/9805201}}].

\bibitem{SupernovaCosmologyProject:1996grv}
{\scshape Supernova Cosmology Project} collaboration, \emph{{Measurements of
  the cosmological parameters Omega and Lambda from the first 7 supernovae at
  z\ensuremath{>}=0.35}},
  \href{https://doi.org/10.1086/304265}{\emph{Astrophys. J.} {\bfseries 483}
  (1997) 565} [\href{https://arxiv.org/abs/astro-ph/9608192}{{\ttfamily
  astro-ph/9608192}}].

\bibitem{Planck:2018vyg}
{\scshape Planck} collaboration, \emph{{Planck 2018 results. VI. Cosmological
  parameters}},
  \href{https://doi.org/10.1051/0004-6361/201833910}{\emph{Astron. Astrophys.}
  {\bfseries 641} (2020) A6}
  [\href{https://arxiv.org/abs/1807.06209}{{\ttfamily 1807.06209}}].

\bibitem{Kachru:2003aw}
S.~Kachru, R.~Kallosh, A.D.~Linde and S.P.~Trivedi, \emph{{De Sitter vacua in
  string theory}},
  \href{https://doi.org/10.1103/PhysRevD.68.046005}{\emph{Phys. Rev. D}
  {\bfseries 68} (2003) 046005}
  [\href{https://arxiv.org/abs/hep-th/0301240}{{\ttfamily hep-th/0301240}}].

\bibitem{Balasubramanian:2005zx}
V.~Balasubramanian, P.~Berglund, J.P.~Conlon and F.~Quevedo, \emph{{Systematics
  of moduli stabilisation in Calabi-Yau flux compactifications}},
  \href{https://doi.org/10.1088/1126-6708/2005/03/007}{\emph{JHEP} {\bfseries
  03} (2005) 007} [\href{https://arxiv.org/abs/hep-th/0502058}{{\ttfamily
  hep-th/0502058}}].

\bibitem{Danielsson:2018ztv}
U.H.~Danielsson and T.~Van~Riet, \emph{{What if string theory has no de Sitter
  vacua?}}, \href{https://doi.org/10.1142/S0218271818300070}{\emph{Int. J. Mod.
  Phys. D} {\bfseries 27} (2018) 1830007}
  [\href{https://arxiv.org/abs/1804.01120}{{\ttfamily 1804.01120}}].

\bibitem{Gao:2020xqh}
X.~Gao, A.~Hebecker and D.~Junghans, \emph{{Control issues of KKLT}},
  \href{https://doi.org/10.1002/prop.202000089}{\emph{Fortsch. Phys.}
  {\bfseries 68} (2020) 2000089}
  [\href{https://arxiv.org/abs/2009.03914}{{\ttfamily 2009.03914}}].

\bibitem{Lust:2022lfc}
S.~L\"ust, C.~Vafa, M.~Wiesner and K.~Xu, \emph{{Holography and the KKLT
  scenario}}, \href{https://doi.org/10.1007/JHEP10(2022)188}{\emph{JHEP}
  {\bfseries 10} (2022) 188}
  [\href{https://arxiv.org/abs/2204.07171}{{\ttfamily 2204.07171}}].

\bibitem{Junghans:2022exo}
D.~Junghans, \emph{{LVS de Sitter vacua are probably in the swampland}},
  \href{https://doi.org/10.1016/j.nuclphysb.2023.116179}{\emph{Nucl. Phys. B}
  {\bfseries 990} (2023) 116179}
  [\href{https://arxiv.org/abs/2201.03572}{{\ttfamily 2201.03572}}].

\bibitem{Cicoli:2023opf}
M.~Cicoli, J.P.~Conlon, A.~Maharana, S.~Parameswaran, F.~Quevedo and I.~Zavala,
  \emph{{String Cosmology: from the Early Universe to Today}},
  \href{https://arxiv.org/abs/2303.04819}{{\ttfamily 2303.04819}}.

\bibitem{Hellerman:2001yi}
S.~Hellerman, N.~Kaloper and L.~Susskind, \emph{{String theory and
  quintessence}},
  \href{https://doi.org/10.1088/1126-6708/2001/06/003}{\emph{JHEP} {\bfseries
  06} (2001) 003} [\href{https://arxiv.org/abs/hep-th/0104180}{{\ttfamily
  hep-th/0104180}}].

\bibitem{Fischler:2001yj}
W.~Fischler, A.~Kashani-Poor, R.~McNees and S.~Paban, \emph{{The Acceleration
  of the universe, a challenge for string theory}},
  \href{https://doi.org/10.1088/1126-6708/2001/07/003}{\emph{JHEP} {\bfseries
  07} (2001) 003} [\href{https://arxiv.org/abs/hep-th/0104181}{{\ttfamily
  hep-th/0104181}}].

\bibitem{Kaloper:2008qs}
N.~Kaloper and L.~Sorbo, \emph{{Where in the String Landscape is
  Quintessence}}, \href{https://doi.org/10.1103/PhysRevD.79.043528}{\emph{Phys.
  Rev. D} {\bfseries 79} (2009) 043528}
  [\href{https://arxiv.org/abs/0810.5346}{{\ttfamily 0810.5346}}].

\bibitem{Cicoli:2012tz}
M.~Cicoli, F.G.~Pedro and G.~Tasinato, \emph{{Natural Quintessence in String
  Theory}}, \href{https://doi.org/10.1088/1475-7516/2012/07/044}{\emph{JCAP}
  {\bfseries 07} (2012) 044} [\href{https://arxiv.org/abs/1203.6655}{{\ttfamily
  1203.6655}}].

\bibitem{Cicoli:2018kdo}
M.~Cicoli, S.~De~Alwis, A.~Maharana, F.~Muia and F.~Quevedo, \emph{{De Sitter
  vs Quintessence in String Theory}},
  \href{https://doi.org/10.1002/prop.201800079}{\emph{Fortsch. Phys.}
  {\bfseries 67} (2019) 1800079}
  [\href{https://arxiv.org/abs/1808.08967}{{\ttfamily 1808.08967}}].

\bibitem{Hebecker:2019csg}
A.~Hebecker, T.~Skrzypek and M.~Wittner, \emph{{The $F$-term Problem and other
  Challenges of Stringy Quintessence}},
  \href{https://doi.org/10.1007/JHEP11(2019)134}{\emph{JHEP} {\bfseries 11}
  (2019) 134} [\href{https://arxiv.org/abs/1909.08625}{{\ttfamily
  1909.08625}}].

\bibitem{ValeixoBento:2020ujr}
B.~Valeixo~Bento, D.~Chakraborty, S.L.~Parameswaran and I.~Zavala, \emph{{Dark
  Energy in String Theory}},
  \href{https://doi.org/10.22323/1.376.0123}{\emph{PoS} {\bfseries CORFU2019}
  (2020) 123} [\href{https://arxiv.org/abs/2005.10168}{{\ttfamily
  2005.10168}}].

\bibitem{Brinkmann:2022oxy}
M.~Brinkmann, M.~Cicoli, G.~Dibitetto and F.G.~Pedro, \emph{{Stringy multifield
  quintessence and the Swampland}},
  \href{https://doi.org/10.1007/JHEP11(2022)044}{\emph{JHEP} {\bfseries 11}
  (2022) 044} [\href{https://arxiv.org/abs/2206.10649}{{\ttfamily
  2206.10649}}].

\bibitem{Vafa:2005ui}
C.~Vafa, \emph{{The String landscape and the swampland}},
  \href{https://arxiv.org/abs/hep-th/0509212}{{\ttfamily hep-th/0509212}}.

\bibitem{vanBeest:2021lhn}
M.~van Beest, J.~Calder\'on-Infante, D.~Mirfendereski and I.~Valenzuela,
  \emph{{Lectures on the Swampland Program in String Compactifications}},
  \href{https://doi.org/10.1016/j.physrep.2022.09.002}{\emph{Phys. Rept.}
  {\bfseries 989} (2022) 1} [\href{https://arxiv.org/abs/2102.01111}{{\ttfamily
  2102.01111}}].

\bibitem{Agmon:2022thq}
N.B.~Agmon, A.~Bedroya, M.J.~Kang and C.~Vafa, \emph{{Lectures on the string
  landscape and the Swampland}},
  \href{https://arxiv.org/abs/2212.06187}{{\ttfamily 2212.06187}}.

\bibitem{Palti:2019pca}
E.~Palti, \emph{{The Swampland: Introduction and Review}},
  \href{https://doi.org/10.1002/prop.201900037}{\emph{Fortsch. Phys.}
  {\bfseries 67} (2019) 1900037}
  [\href{https://arxiv.org/abs/1903.06239}{{\ttfamily 1903.06239}}].

\bibitem{Grana:2021zvf}
M.~Gra\~na and A.~Herr\'aez, \emph{{The Swampland Conjectures: A Bridge from
  Quantum Gravity to Particle Physics}},
  \href{https://doi.org/10.3390/universe7080273}{\emph{Universe} {\bfseries 7}
  (2021) 273} [\href{https://arxiv.org/abs/2107.00087}{{\ttfamily
  2107.00087}}].

\bibitem{Ooguri:2006in}
H.~Ooguri and C.~Vafa, \emph{{On the Geometry of the String Landscape and the
  Swampland}},
  \href{https://doi.org/10.1016/j.nuclphysb.2006.10.033}{\emph{Nucl. Phys. B}
  {\bfseries 766} (2007) 21}
  [\href{https://arxiv.org/abs/hep-th/0605264}{{\ttfamily hep-th/0605264}}].

\bibitem{Obied:2018sgi}
G.~Obied, H.~Ooguri, L.~Spodyneiko and C.~Vafa, \emph{{De Sitter Space and the
  Swampland}},  \href{https://arxiv.org/abs/1806.08362}{{\ttfamily
  1806.08362}}.

\bibitem{Banlaki:2018ayh}
A.~Banlaki, A.~Chowdhury, C.~Roupec and T.~Wrase, \emph{{Scaling limits of dS
  vacua and the swampland}},
  \href{https://doi.org/10.1007/JHEP03(2019)065}{\emph{JHEP} {\bfseries 03}
  (2019) 065} [\href{https://arxiv.org/abs/1811.07880}{{\ttfamily
  1811.07880}}].

\bibitem{Junghans:2018gdb}
D.~Junghans, \emph{{Weakly Coupled de Sitter Vacua with Fluxes and the
  Swampland}}, \href{https://doi.org/10.1007/JHEP03(2019)150}{\emph{JHEP}
  {\bfseries 03} (2019) 150}
  [\href{https://arxiv.org/abs/1811.06990}{{\ttfamily 1811.06990}}].

\bibitem{Ooguri:2018wrx}
H.~Ooguri, E.~Palti, G.~Shiu and C.~Vafa, \emph{{Distance and de Sitter
  Conjectures on the Swampland}},
  \href{https://doi.org/10.1016/j.physletb.2018.11.018}{\emph{Phys. Lett. B}
  {\bfseries 788} (2019) 180}
  [\href{https://arxiv.org/abs/1810.05506}{{\ttfamily 1810.05506}}].

\bibitem{Hebecker:2018vxz}
A.~Hebecker and T.~Wrase, \emph{{The Asymptotic dS Swampland Conjecture - a
  Simplified Derivation and a Potential Loophole}},
  \href{https://doi.org/10.1002/prop.201800097}{\emph{Fortsch. Phys.}
  {\bfseries 67} (2019) 1800097}
  [\href{https://arxiv.org/abs/1810.08182}{{\ttfamily 1810.08182}}].

\bibitem{Lust:2019zwm}
D.~L\"ust, E.~Palti and C.~Vafa, \emph{{AdS and the Swampland}},
  \href{https://doi.org/10.1016/j.physletb.2019.134867}{\emph{Phys. Lett. B}
  {\bfseries 797} (2019) 134867}
  [\href{https://arxiv.org/abs/1906.05225}{{\ttfamily 1906.05225}}].

\bibitem{Hertzberg:2007wc}
M.P.~Hertzberg, S.~Kachru, W.~Taylor and M.~Tegmark, \emph{{Inflationary
  Constraints on Type IIA String Theory}},
  \href{https://doi.org/10.1088/1126-6708/2007/12/095}{\emph{JHEP} {\bfseries
  12} (2007) 095} [\href{https://arxiv.org/abs/0711.2512}{{\ttfamily
  0711.2512}}].

\bibitem{Haque:2008jz}
S.S.~Haque, G.~Shiu, B.~Underwood and T.~Van~Riet, \emph{{Minimal simple de
  Sitter solutions}},
  \href{https://doi.org/10.1103/PhysRevD.79.086005}{\emph{Phys. Rev. D}
  {\bfseries 79} (2009) 086005}
  [\href{https://arxiv.org/abs/0810.5328}{{\ttfamily 0810.5328}}].

\bibitem{Flauger:2008ad}
R.~Flauger, S.~Paban, D.~Robbins and T.~Wrase, \emph{{Searching for slow-roll
  moduli inflation in massive type IIA supergravity with metric fluxes}},
  \href{https://doi.org/10.1103/PhysRevD.79.086011}{\emph{Phys. Rev. D}
  {\bfseries 79} (2009) 086011}
  [\href{https://arxiv.org/abs/0812.3886}{{\ttfamily 0812.3886}}].

\bibitem{Caviezel:2009tu}
C.~Caviezel, T.~Wrase and M.~Zagermann, \emph{{Moduli Stabilization and
  Cosmology of Type IIB on SU(2)-Structure Orientifolds}},
  \href{https://doi.org/10.1007/JHEP04(2010)011}{\emph{JHEP} {\bfseries 04}
  (2010) 011} [\href{https://arxiv.org/abs/0912.3287}{{\ttfamily 0912.3287}}].

\bibitem{Wrase:2010ew}
T.~Wrase and M.~Zagermann, \emph{{On Classical de Sitter Vacua in String
  Theory}}, \href{https://doi.org/10.1002/prop.201000053}{\emph{Fortsch. Phys.}
  {\bfseries 58} (2010) 906} [\href{https://arxiv.org/abs/1003.0029}{{\ttfamily
  1003.0029}}].

\bibitem{Shiu:2011zt}
G.~Shiu and Y.~Sumitomo, \emph{{Stability Constraints on Classical de Sitter
  Vacua}}, \href{https://doi.org/10.1007/JHEP09(2011)052}{\emph{JHEP}
  {\bfseries 09} (2011) 052} [\href{https://arxiv.org/abs/1107.2925}{{\ttfamily
  1107.2925}}].

\bibitem{Andriot:2016xvq}
D.~Andriot and J.~Bl\r{a}b\"ack, \emph{{Refining the boundaries of the
  classical de Sitter landscape}},
  \href{https://doi.org/10.1007/JHEP03(2017)102}{\emph{JHEP} {\bfseries 03}
  (2017) 102} [\href{https://arxiv.org/abs/1609.00385}{{\ttfamily
  1609.00385}}].

\bibitem{Garg:2018reu}
S.K.~Garg and C.~Krishnan, \emph{{Bounds on Slow Roll and the de Sitter
  Swampland}}, \href{https://doi.org/10.1007/JHEP11(2019)075}{\emph{JHEP}
  {\bfseries 11} (2019) 075}
  [\href{https://arxiv.org/abs/1807.05193}{{\ttfamily 1807.05193}}].

\bibitem{Andriot:2018ept}
D.~Andriot, \emph{{New constraints on classical de Sitter: flirting with the
  swampland}}, \href{https://doi.org/10.1002/prop.201800103}{\emph{Fortsch.
  Phys.} {\bfseries 67} (2019) 1800103}
  [\href{https://arxiv.org/abs/1807.09698}{{\ttfamily 1807.09698}}].

\bibitem{Garg:2018zdg}
S.K.~Garg, C.~Krishnan and M.~Zaid~Zaz, \emph{{Bounds on Slow Roll at the
  Boundary of the Landscape}},
  \href{https://doi.org/10.1007/JHEP03(2019)029}{\emph{JHEP} {\bfseries 03}
  (2019) 029} [\href{https://arxiv.org/abs/1810.09406}{{\ttfamily
  1810.09406}}].

\bibitem{Andriot:2022xjh}
D.~Andriot and L.~Horer, \emph{{(Quasi-) de Sitter solutions across dimensions
  and the TCC bound}},
  \href{https://doi.org/10.1007/JHEP01(2023)020}{\emph{JHEP} {\bfseries 01}
  (2023) 020} [\href{https://arxiv.org/abs/2208.14462}{{\ttfamily
  2208.14462}}].

\bibitem{Andriot:2023isc}
D.~Andriot, \emph{{Bumping into the species scale with the scalar potential}},
  \href{https://arxiv.org/abs/2305.07480}{{\ttfamily 2305.07480}}.

\bibitem{Cicoli:2021fsd}
M.~Cicoli, F.~Cunillera, A.~Padilla and F.G.~Pedro, \emph{{Quintessence and the
  Swampland: The Parametrically Controlled Regime of Moduli Space}},
  \href{https://doi.org/10.1002/prop.202200009}{\emph{Fortsch. Phys.}
  {\bfseries 70} (2022) 2200009}
  [\href{https://arxiv.org/abs/2112.10779}{{\ttfamily 2112.10779}}].

\bibitem{Dine:1985he}
M.~Dine and N.~Seiberg, \emph{{Is the Superstring Weakly Coupled?}},
  \href{https://doi.org/10.1016/0370-2693(85)90927-X}{\emph{Phys. Lett. B}
  {\bfseries 162} (1985) 299}.

\bibitem{Rudelius:2021azq}
T.~Rudelius, \emph{{Asymptotic observables and the swampland}},
  \href{https://doi.org/10.1103/PhysRevD.104.126023}{\emph{Phys. Rev. D}
  {\bfseries 104} (2021) 126023}
  [\href{https://arxiv.org/abs/2106.09026}{{\ttfamily 2106.09026}}].

\bibitem{Rudelius:2021oaz}
T.~Rudelius, \emph{{Dimensional reduction and (Anti) de Sitter bounds}},
  \href{https://doi.org/10.1007/JHEP08(2021)041}{\emph{JHEP} {\bfseries 08}
  (2021) 041} [\href{https://arxiv.org/abs/2101.11617}{{\ttfamily
  2101.11617}}].

\bibitem{Etheredge:2022opl}
M.~Etheredge, B.~Heidenreich, S.~Kaya, Y.~Qiu and T.~Rudelius,
  \emph{{Sharpening the Distance Conjecture in diverse dimensions}},
  \href{https://doi.org/10.1007/JHEP12(2022)114}{\emph{JHEP} {\bfseries 12}
  (2022) 114} [\href{https://arxiv.org/abs/2206.04063}{{\ttfamily
  2206.04063}}].

\bibitem{Rudelius:2022gbz}
T.~Rudelius, \emph{{Asymptotic scalar field cosmology in string theory}},
  \href{https://doi.org/10.1007/JHEP10(2022)018}{\emph{JHEP} {\bfseries 10}
  (2022) 018} [\href{https://arxiv.org/abs/2208.08989}{{\ttfamily
  2208.08989}}].

\bibitem{Bedroya:2019snp}
A.~Bedroya and C.~Vafa, \emph{{Trans-Planckian Censorship and the Swampland}},
  \href{https://doi.org/10.1007/JHEP09(2020)123}{\emph{JHEP} {\bfseries 09}
  (2020) 123} [\href{https://arxiv.org/abs/1909.11063}{{\ttfamily
  1909.11063}}].

\bibitem{Marconnet:2022fmx}
P.~Marconnet and D.~Tsimpis, \emph{{Universal accelerating cosmologies from 10d
  supergravity}}, \href{https://doi.org/10.1007/JHEP01(2023)033}{\emph{JHEP}
  {\bfseries 01} (2023) 033}
  [\href{https://arxiv.org/abs/2210.10813}{{\ttfamily 2210.10813}}].

\bibitem{Bedroya:2022tbh}
A.~Bedroya, \emph{{Holographic origin of TCC and the Distance Conjecture}},
  \href{https://arxiv.org/abs/2211.09128}{{\ttfamily 2211.09128}}.

\bibitem{Apers:2022cyl}
F.~Apers, J.P.~Conlon, M.~Mosny and F.~Revello, \emph{{Kination, Meet Kasner:
  On The Asymptotic Cosmology of String Compactifications}},
  \href{https://arxiv.org/abs/2212.10293}{{\ttfamily 2212.10293}}.

\bibitem{Shiu:2023nph}
G.~Shiu, F.~Tonioni and H.V.~Tran, \emph{{Accelerating universe at the end of
  time}},  \href{https://arxiv.org/abs/2303.03418}{{\ttfamily 2303.03418}}.

\bibitem{Shiu:2023rxt}
G.~Shiu, F.~Tonioni and H.V.~Tran, \emph{{Late-time attractors and cosmic
  acceleration}},  \href{https://arxiv.org/abs/2306.07327}{{\ttfamily
  2306.07327}}.

\bibitem{Calderon-Infante:2022nxb}
J.~Calder\'on-Infante, I.~Ruiz and I.~Valenzuela, \emph{{Asymptotic Accelerated
  Expansion in String Theory and the Swampland}},
  \href{https://arxiv.org/abs/2209.11821}{{\ttfamily 2209.11821}}.

\bibitem{Becker:2006ks}
K.~Becker, M.~Becker, C.~Vafa and J.~Walcher, \emph{{Moduli Stabilization in
  Non-Geometric Backgrounds}},
  \href{https://doi.org/10.1016/j.nuclphysb.2007.01.034}{\emph{Nucl. Phys. B}
  {\bfseries 770} (2007) 1}
  [\href{https://arxiv.org/abs/hep-th/0611001}{{\ttfamily hep-th/0611001}}].

\bibitem{Becker:2007dn}
K.~Becker, M.~Becker and J.~Walcher, \emph{{Runaway in the Landscape}},
  \href{https://doi.org/10.1103/PhysRevD.76.106002}{\emph{Phys. Rev. D}
  {\bfseries 76} (2007) 106002}
  [\href{https://arxiv.org/abs/0706.0514}{{\ttfamily 0706.0514}}].

\bibitem{Ishiguro:2021csu}
K.~Ishiguro and H.~Otsuka, \emph{{Sharpening the boundaries between flux
  landscape and swampland by tadpole charge}},
  \href{https://doi.org/10.1007/JHEP12(2021)017}{\emph{JHEP} {\bfseries 12}
  (2021) 017} [\href{https://arxiv.org/abs/2104.15030}{{\ttfamily
  2104.15030}}].

\bibitem{Bardzell:2022jfh}
J.~Bardzell, E.~Gonzalo, M.~Rajaguru, D.~Smith and T.~Wrase, \emph{{Type IIB
  flux compactifications with h$^{1,1}$ = 0}},
  \href{https://doi.org/10.1007/JHEP06(2022)166}{\emph{JHEP} {\bfseries 06}
  (2022) 166} [\href{https://arxiv.org/abs/2203.15818}{{\ttfamily
  2203.15818}}].

\bibitem{Becker:2022hse}
K.~Becker, E.~Gonzalo, J.~Walcher and T.~Wrase, \emph{{Fluxes, vacua, and
  tadpoles meet Landau-Ginzburg and Fermat}},
  \href{https://doi.org/10.1007/JHEP12(2022)083}{\emph{JHEP} {\bfseries 12}
  (2022) 083} [\href{https://arxiv.org/abs/2210.03706}{{\ttfamily
  2210.03706}}].

\bibitem{Klaewer:2016kiy}
D.~Klaewer and E.~Palti, \emph{{Super-Planckian Spatial Field Variations and
  Quantum Gravity}}, \href{https://doi.org/10.1007/JHEP01(2017)088}{\emph{JHEP}
  {\bfseries 01} (2017) 088}
  [\href{https://arxiv.org/abs/1610.00010}{{\ttfamily 1610.00010}}].

\bibitem{Lee:2019wij}
S.-J.~Lee, W.~Lerche and T.~Weigand, \emph{{Emergent strings from infinite
  distance limits}}, \href{https://doi.org/10.1007/JHEP02(2022)190}{\emph{JHEP}
  {\bfseries 02} (2022) 190}
  [\href{https://arxiv.org/abs/1910.01135}{{\ttfamily 1910.01135}}].

\bibitem{Lee:2019xtm}
S.-J.~Lee, W.~Lerche and T.~Weigand, \emph{{Emergent strings, duality and weak
  coupling limits for two-form fields}},
  \href{https://doi.org/10.1007/JHEP02(2022)096}{\emph{JHEP} {\bfseries 02}
  (2022) 096} [\href{https://arxiv.org/abs/1904.06344}{{\ttfamily
  1904.06344}}].

\bibitem{DeWolfe:2005uu}
O.~DeWolfe, A.~Giryavets, S.~Kachru and W.~Taylor, \emph{{Type IIA moduli
  stabilization}},
  \href{https://doi.org/10.1088/1126-6708/2005/07/066}{\emph{JHEP} {\bfseries
  07} (2005) 066} [\href{https://arxiv.org/abs/hep-th/0505160}{{\ttfamily
  hep-th/0505160}}].

\bibitem{Gukov:1999ya}
S.~Gukov, C.~Vafa and E.~Witten, \emph{{CFT's from Calabi-Yau four folds}},
  \href{https://doi.org/10.1016/S0550-3213(00)00373-4}{\emph{Nucl. Phys. B}
  {\bfseries 584} (2000) 69}
  [\href{https://arxiv.org/abs/hep-th/9906070}{{\ttfamily hep-th/9906070}}].

\bibitem{Hamada:2021yxy}
Y.~Hamada, M.~Montero, C.~Vafa and I.~Valenzuela, \emph{{Finiteness and the
  swampland}}, \href{https://doi.org/10.1088/1751-8121/ac6404}{\emph{J. Phys.
  A} {\bfseries 55} (2022) 224005}
  [\href{https://arxiv.org/abs/2111.00015}{{\ttfamily 2111.00015}}].

\bibitem{vandeHeisteeg:2023ubh}
D.~van~de Heisteeg, C.~Vafa and M.~Wiesner, \emph{{Bounds on Species Scale and
  the Distance Conjecture}},
  \href{https://arxiv.org/abs/2303.13580}{{\ttfamily 2303.13580}}.

\bibitem{vandeHeisteeg:2023uxj}
D.~van~de Heisteeg, C.~Vafa, M.~Wiesner and D.H.~Wu, \emph{{Bounds on Field
  Range for Slowly Varying Positive Potentials}},
  \href{https://arxiv.org/abs/2305.07701}{{\ttfamily 2305.07701}}.

\bibitem{Dias:2018pgj}
M.~Dias, J.~Frazer, A.~Retolaza, M.~Scalisi and A.~Westphal, \emph{{Pole
  N-flation}}, \href{https://doi.org/10.1007/JHEP02(2019)120}{\emph{JHEP}
  {\bfseries 02} (2019) 120}
  [\href{https://arxiv.org/abs/1805.02659}{{\ttfamily 1805.02659}}].

\bibitem{Andriot:2022brg}
D.~Andriot, L.~Horer and G.~Tringas, \emph{{Negative scalar potentials and the
  swampland: an Anti-Trans-Planckian Censorship Conjecture}},
  \href{https://doi.org/10.1007/JHEP04(2023)139}{\emph{JHEP} {\bfseries 04}
  (2023) 139} [\href{https://arxiv.org/abs/2212.04517}{{\ttfamily
  2212.04517}}].

\bibitem{Andriot:2025gyr}
D.~Andriot, M.~Rajaguru and G.~Tringas, \emph{{Single versus multifield scalar
  potentials from string theory}},
  \href{https://arxiv.org/abs/2501.17775}{{\ttfamily 2501.17775}}.

\bibitem{Andriot:2020lea}
D.~Andriot, N.~Cribiori and D.~Erkinger, \emph{{The web of swampland
  conjectures and the TCC bound}},
  \href{https://doi.org/10.1007/JHEP07(2020)162}{\emph{JHEP} {\bfseries 07}
  (2020) 162} [\href{https://arxiv.org/abs/2004.00030}{{\ttfamily
  2004.00030}}].

\bibitem{Bedroya:2020rmd}
A.~Bedroya, \emph{{de Sitter Complementarity, TCC, and the Swampland}},
  \href{https://doi.org/10.31526/lhep.2021.187}{\emph{LHEP} {\bfseries 2021}
  (2021) 187} [\href{https://arxiv.org/abs/2010.09760}{{\ttfamily
  2010.09760}}].

\bibitem{Montero:2022prj}
M.~Montero, C.~Vafa and I.~Valenzuela, \emph{{The dark dimension and the
  Swampland}}, \href{https://doi.org/10.1007/JHEP02(2023)022}{\emph{JHEP}
  {\bfseries 02} (2023) 022}
  [\href{https://arxiv.org/abs/2205.12293}{{\ttfamily 2205.12293}}].

\end{thebibliography}\endgroup

\end{document}